\pdfoutput=1

\documentclass[12pt,a4paper]{article}

\usepackage{ifthen}
\newboolean{pdflatex}
\setboolean{pdflatex}{true}

\newboolean{articletitles}
\setboolean{articletitles}{true}

\newboolean{uprightparticles}
\setboolean{uprightparticles}{false}

\newboolean{inbibliography}
\setboolean{inbibliography}{false}

\usepackage[top=1in, bottom=1.25in, left=1in, right=1in]{geometry}
\usepackage{microtype}
\usepackage{lineno}
\usepackage{xspace}
\usepackage{caption}

\usepackage{graphicx}
\usepackage{color}
\usepackage{colortbl}
\graphicspath{{./figs/}}
\usepackage{amsmath}
\usepackage{amssymb}
\usepackage{amsfonts}
\usepackage{upgreek}
\newcommand*\patchAmsMathEnvironmentForLineno[1]{%
\expandafter\let\csname old#1\expandafter\endcsname\csname #1\endcsname
\expandafter\let\csname oldend#1\expandafter\endcsname\csname
end#1\endcsname
 \renewenvironment{#1}%
   {\linenomath\csname old#1\endcsname}%
   {\csname oldend#1\endcsname\endlinenomath}%
}
\newcommand*\patchBothAmsMathEnvironmentsForLineno[1]{%
  \patchAmsMathEnvironmentForLineno{#1}%
  \patchAmsMathEnvironmentForLineno{#1*}%
}
\AtBeginDocument{%
\patchBothAmsMathEnvironmentsForLineno{equation}%
\patchBothAmsMathEnvironmentsForLineno{align}%
\patchBothAmsMathEnvironmentsForLineno{flalign}%
\patchBothAmsMathEnvironmentsForLineno{alignat}%
\patchBothAmsMathEnvironmentsForLineno{gather}%
\patchBothAmsMathEnvironmentsForLineno{multline}%
\patchBothAmsMathEnvironmentsForLineno{eqnarray}%
}
\usepackage{hyperref}
\usepackage[all]{hypcap}

\usepackage{xspace} 
\usepackage{upgreek}

\def\lhcb {\mbox{LHCb}\xspace}

\def\MagUp {\mbox{\em Mag\kern -0.05em Up}\xspace}

\ifthenelse{\boolean{uprightparticles}}%
{

 \def\Pmu         {\ensuremath{\upmu}\xspace}                 
 \def\Pnu         {\ensuremath{\upnu}\xspace}                 
                  
 \def\Ppi         {\ensuremath{\uppi}\xspace}                 
                  
 \def\Prho        {\ensuremath{\uprho}\xspace}

 \def\Pphi        {\ensuremath{\upphi}\xspace}

 \def\Ppsi        {\ensuremath{\uppsi}\xspace}

 \def\PDelta      {\ensuremath{\Delta}\xspace}                 
 \def\PXi      {\ensuremath{\Xi}\xspace}                 
 \def\PLambda      {\ensuremath{\Lambda}\xspace}                 
 \def\PSigma      {\ensuremath{\Sigma}\xspace}                 
 \def\POmega      {\ensuremath{\Omega}\xspace}                 
 \def\PUpsilon      {\ensuremath{\Upsilon}\xspace}                 
 
 \def\PB      {\ensuremath{\mathrm{B}}\xspace}                 
                  
 \def\PD      {\ensuremath{\mathrm{D}}\xspace}

 \def\PJ      {\ensuremath{\mathrm{J}}\xspace}                 
 \def\PK      {\ensuremath{\mathrm{K}}\xspace}

 \def\Pb      {\ensuremath{\mathrm{b}}\xspace}                 
 \def\Pc      {\ensuremath{\mathrm{c}}\xspace}                 
 \def\Pd      {\ensuremath{\mathrm{d}}\xspace}

 \def\Pi      {\ensuremath{\mathrm{i}}\xspace}

 \def\Pp      {\ensuremath{\mathrm{p}}\xspace}

 \def\Ps      {\ensuremath{\mathrm{s}}\xspace}

}
{

 \def\Pmu         {\ensuremath{\mu}\xspace}                 
 \def\Pnu         {\ensuremath{\nu}\xspace}                 
                  
 \def\Ppi         {\ensuremath{\pi}\xspace}                 
                  
 \def\Prho        {\ensuremath{\rho}\xspace}

 \def\Pphi        {\ensuremath{\phi}\xspace}

 \def\Ppsi        {\ensuremath{\psi}\xspace}                 
                  
 \mathchardef\PDelta="7101
 \mathchardef\PXi="7104
 \mathchardef\PLambda="7103
 \mathchardef\PSigma="7106
 \mathchardef\POmega="710A
 \mathchardef\PUpsilon="7107
                  
 \def\PB      {\ensuremath{B}\xspace}                 
                  
 \def\PD      {\ensuremath{D}\xspace}

 \def\PJ      {\ensuremath{J}\xspace}                 
 \def\PK      {\ensuremath{K}\xspace}

 \def\Pb      {\ensuremath{b}\xspace}                 
 \def\Pc      {\ensuremath{c}\xspace}                 
 \def\Pd      {\ensuremath{d}\xspace}

 \def\Pi      {\ensuremath{i}\xspace}

 \def\Pp      {\ensuremath{p}\xspace}

 \def\Ps      {\ensuremath{s}\xspace}

}

\makeatletter
\ifcase \@ptsize \relax
  \newcommand{\miniscule}{\@setfontsize\miniscule{4}{5}}
\or
  \newcommand{\miniscule}{\@setfontsize\miniscule{5}{6}}
\or
  \newcommand{\miniscule}{\@setfontsize\miniscule{5}{6}}
\fi
\makeatother

\DeclareRobustCommand{\optbar}[1]{\shortstack{{\miniscule (\rule[.5ex]{1.25em}{.18mm})}
  \\ [-.7ex] $#1$}}

\def\mun        {{\ensuremath{\Pmu^-}}\xspace} 

\def\neub       {{\ensuremath{\overline{\Pnu}}}\xspace}

\def\neumb      {{\ensuremath{\neub_\mu}}\xspace}

\def\dquark    {{\ensuremath{\Pd}}\xspace}
\def\dquarkbar {{\ensuremath{\overline \dquark}}\xspace}

\def\squark    {{\ensuremath{\Ps}}\xspace}
\def\squarkbar {{\ensuremath{\overline \squark}}\xspace}

\def\cquark    {{\ensuremath{\Pc}}\xspace}
\def\cquarkbar {{\ensuremath{\overline \cquark}}\xspace}

\def\bquark    {{\ensuremath{\Pb}}\xspace}
\def\bquarkbar {{\ensuremath{\overline \bquark}}\xspace}
\def\bbbar     {{\ensuremath{\bquark\bquarkbar}}\xspace}

\def\pion   {{\ensuremath{\Ppi}}\xspace}

\def\pip    {{\ensuremath{\pion^+}}\xspace}
\def\pim    {{\ensuremath{\pion^-}}\xspace}
\def\pipm   {{\ensuremath{\pion^\pm}}\xspace}
\def\pimp   {{\ensuremath{\pion^\mp}}\xspace}

\def\rhomeson {{\ensuremath{\Prho}}\xspace}
\def\rhoz     {{\ensuremath{\rhomeson^0}}\xspace}

\def\kaon    {{\ensuremath{\PK}}\xspace}
  \def\Kbar    {{\kern 0.2em\overline{\kern -0.2em \PK}{}}\xspace}

\def\KorKbar    {\kern 0.18em\optbar{\kern -0.18em K}{}\xspace}

\def\Kp      {{\ensuremath{\kaon^+}}\xspace}
\def\Km      {{\ensuremath{\kaon^-}}\xspace}
\def\Kpm     {{\ensuremath{\kaon^\pm}}\xspace}

\def\Kstarz  {{\ensuremath{\kaon^{*0}}}\xspace}
\def\Kstarzb {{\ensuremath{\Kbar{}^{*0}}}\xspace}

\newcommand{\phiz}{\ensuremath{\Pphi}\xspace}

  \def\Dbar    {{\kern 0.2em\overline{\kern -0.2em \PD}{}}\xspace}
\def\D       {{\ensuremath{\PD}}\xspace}

\def\DorDbar    {\kern 0.18em\optbar{\kern -0.18em D}{}\xspace}
\def\Dz      {{\ensuremath{\D^0}}\xspace}

\def\Dstarp  {{\ensuremath{\D^{*+}}}\xspace}

\def\Dsm     {{\ensuremath{\D^-_\squark}}\xspace}

\def\B       {{\ensuremath{\PB}}\xspace}
\def\Bbar    {{\ensuremath{\kern 0.18em\overline{\kern -0.18em \PB}{}}}\xspace}

\def\BorBbar    {\kern 0.18em\optbar{\kern -0.18em B}{}\xspace}

\def\Bd      {{\ensuremath{\B^0}}\xspace}
\def\Bs      {{\ensuremath{\B^0_\squark}}\xspace}
\def\Bsb     {{\ensuremath{\Bbar{}^0_\squark}}\xspace}

\def\jpsi     {{\ensuremath{{\PJ\mskip -3mu/\mskip -2mu\Ppsi\mskip 2mu}}}\xspace}

  \def\Y#1S{\ensuremath{\PUpsilon{(#1S)}}\xspace}

\def\proton      {{\ensuremath{\Pp}}\xspace}

\def\Lz          {{\ensuremath{\PLambda}}\xspace}
\def\Lbar        {{\ensuremath{\kern 0.1em\overline{\kern -0.1em\PLambda}}}\xspace}
\def\LorLbar    {\kern 0.18em\optbar{\kern -0.18em \PLambda}{}\xspace}

\def\Lb      {{\ensuremath{\Lz^0_\bquark}}\xspace}

\def\Lc      {{\ensuremath{\Lz^+_\cquark}}\xspace}

\def\to                 {\ensuremath{\rightarrow}\xspace}

\def\CP                {{\ensuremath{C\!P}}\xspace}

\newcommand{\phis}{{\ensuremath{\phi_{\squark}}}\xspace}

\def\AT#1     {\ensuremath{A_{\mathrm{T}}^{#1}}\xspace}           

\def\C#1      {\ensuremath{\mathcal{C}_{#1}}\xspace}                       
\def\Cp#1     {\ensuremath{\mathcal{C}_{#1}^{'}}\xspace}                    
\def\Ceff#1   {\ensuremath{\mathcal{C}_{#1}^{\mathrm{(eff)}}}\xspace}        
\def\Cpeff#1  {\ensuremath{\mathcal{C}_{#1}^{'\mathrm{(eff)}}}\xspace}       
\def\Ope#1    {\ensuremath{\mathcal{O}_{#1}}\xspace}                       
\def\Opep#1   {\ensuremath{\mathcal{O}_{#1}^{'}}\xspace}                    

\newcommand{\ket}[1]{\ensuremath{|#1\rangle}}              
\newcommand{\braket}[2]{\ensuremath{\langle #1|#2\rangle}} 

\newcommand{\tev}{\ifthenelse{\boolean{inbibliography}}{\ensuremath{~T\kern -0.05em eV}}{\ensuremath{\mathrm{\,Te\kern -0.1em V}}}\xspace}
\newcommand{\gev}{\ensuremath{\mathrm{\,Ge\kern -0.1em V}}\xspace}
\newcommand{\mev}{\ensuremath{\mathrm{\,Me\kern -0.1em V}}\xspace}
\newcommand{\kev}{\ensuremath{\mathrm{\,ke\kern -0.1em V}}\xspace}
\newcommand{\ev}{\ensuremath{\mathrm{\,e\kern -0.1em V}}\xspace}
\newcommand{\gevc}{\ensuremath{{\mathrm{\,Ge\kern -0.1em V\!/}c}}\xspace}
\newcommand{\mevc}{\ensuremath{{\mathrm{\,Me\kern -0.1em V\!/}c}}\xspace}
\newcommand{\gevcc}{\ensuremath{{\mathrm{\,Ge\kern -0.1em V\!/}c^2}}\xspace}
\newcommand{\gevgevcccc}{\ensuremath{{\mathrm{\,Ge\kern -0.1em V^2\!/}c^4}}\xspace}
\newcommand{\mevcc}{\ensuremath{{\mathrm{\,Me\kern -0.1em V\!/}c^2}}\xspace}

\def\mum  {\ensuremath{{\,\upmu\mathrm{m}}}\xspace}

\def\invpb {\ensuremath{\mbox{\,pb}^{-1}}\xspace}

\def\invfb   {\ensuremath{\mbox{\,fb}^{-1}}\xspace}

\def\ms   {\ensuremath{{\mathrm{ \,ms}}}\xspace}

\newcommand{\stat}{\ensuremath{\mathrm{\,(stat)}}\xspace}
\newcommand{\syst}{\ensuremath{\mathrm{\,(syst)}}\xspace}

\newcommand{\chisq}{\ensuremath{\chi^2}\xspace}

\newcommand{\chisqip}{\ensuremath{\chi^2_{\text{IP}}}\xspace}

\def\gsim{{~\raise.15em\hbox{$>$}\kern-.85em
          \lower.35em\hbox{$\sim$}~}\xspace}
\def\lsim{{~\raise.15em\hbox{$<$}\kern-.85em
          \lower.35em\hbox{$\sim$}~}\xspace}

\def\sPlot{\mbox{\em sPlot}\xspace}

\def\ptot       {\mbox{$p$}\xspace}
\def\pt         {\mbox{$p_{\mathrm{ T}}$}\xspace}

\def\rad{\ensuremath{\mathrm{ \,rad}}\xspace}

\def\evtgen     {\mbox{\textsc{EvtGen}}\xspace}

\def\geant      {\mbox{\textsc{Geant4}}\xspace}

\def\photos     {\mbox{\textsc{Photos}}\xspace}

\def\pythia     {\mbox{\textsc{Pythia}}\xspace}

\def\tell1  {TELL1\xspace}
\def\ukl1   {UKL1\xspace}

\usepackage{cite}
\usepackage{mciteplus}
\usepackage{longtable}
\usepackage{rotating}

\usepackage[bottom]{footmisc}

\newcolumntype{x}[1]{>{\centering\let\newline\\\arraybackslash\hspace{0pt}}p{#1}}
\newcolumntype{k}[1]{>{\raggedleft\let\newline\\\arraybackslash\hspace{0pt}}p{#1}}
\def\Kstarzp  {{\ensuremath{\kaon^{*}(892)^{0}}}\xspace}
\def\Kstarzbp {{\ensuremath{\Kbar{}^{*}(892)^{0}}}\xspace}
\def\Kstarzss  {{\ensuremath{\kaon_0^{*}(800)^{0}}}\xspace}
\def\Kstarzs  {{\ensuremath{\kaon_0^{*}(1430)^{0}}}\xspace}
\def\Kstarzbs {{\ensuremath{\Kbar{}_0^{*}(1430)^{0}}}\xspace}
\def\Kstarzd  {{\ensuremath{\kaon_2^{*}(1430)^{0}}}\xspace}
\def\Kstarzbd {{\ensuremath{\Kbar{}_2^{*}(1430)^{0}}}\xspace}
\newcommand{\Kpi}{\ensuremath{\Kp\pim}\xspace}
\newcommand{\Kpib}{\ensuremath{\Km\pip}\xspace}
\newcommand{\Kpipm}{\ensuremath{\Kpm\pimp}\xspace}
\newcommand{\BsKstKst}{\ensuremath{{\Bs\to\Kstarz\Kstarzb}}\xspace}

\newcommand{\BsKstKsts}{\ensuremath{{\Bs\to\Kstarz\Kstarzbs}}\xspace}

\newcommand{\BsKstsKsts}{\ensuremath{{\Bs\to\Kstarzs\Kstarzbs}}\xspace}

\newcommand{\BsKpiKpi}{\ensuremath{{\Bs\to(\Kpi)(\Kpib)}}\xspace}
\newcommand{\BsKpiKpiPS}{\ensuremath{{\Bs\to\Kpi\Kpib}}\xspace}
\newcommand{\BdKstKst}{\ensuremath{{\Bd\to\Kstarz\Kstarzb}}\xspace}
\newcommand{\BdKstRho}{\ensuremath{{\Bd\to\Kstarz\rhoz}}\xspace}
\newcommand{\BdKstPhi}{\ensuremath{{\Bd\to\Kstarz\phiz}}\xspace}
\newcommand{\LbPKpipi}{\ensuremath{{\Lb\to(\proton\Km)(\pip\pim)}}\xspace}
\newcommand{\phisss}{\ensuremath{\phi_{s}^{\squark\squarkbar}}\xspace}
\newcommand{\phisdd}{\ensuremath{\phi_s^{\dquark\dquarkbar}}\xspace}
\newcommand{\phiscc}{\ensuremath{\phi_s^{\cquark\cquarkbar}}\xspace}
\newcommand{\btoddbs}{\ensuremath{\bquark\to\dquark\dquarkbar\squark}\xspace}
\newcommand{\btoccbs}{\ensuremath{\bquark\to\cquark\cquarkbar\squark}\xspace}

\newcommand{\MeVcc}{\ensuremath{\,{\rm MeV}/c^2}\xspace}

\newcommand{\fL}{\ensuremath{f_{\rm L}}\xspace}
\newcommand{\BsH}{\ensuremath{B^{0}_{s,\rm{H}}}\xspace}
\newcommand{\BsL}{\ensuremath{B^{0}_{s,\rm{L}}}\xspace}
\newcommand{\mBH}{\ensuremath{m_{\BsH}}\xspace}
\newcommand{\mBL}{\ensuremath{m_{\BsL}}\xspace}
\newcommand{\GammaBH}{\ensuremath{\Gamma_{\BsH}}\xspace}
\newcommand{\GammaBL}{\ensuremath{\Gamma_{\BsL}}\xspace}
\newcommand{\Deltams}{\ensuremath{\Delta m_{s}}\xspace}
\newcommand{\DeltaGammas}{\ensuremath{\Delta\Gamma_{s}}\xspace}
\renewcommand{\ms}{\ensuremath{m_{s}}\xspace}
\newcommand{\Gammas}{\ensuremath{\Gamma_{s}}\xspace}
\newcommand{\swave}{\ensuremath{S}-wave\xspace}

\newcommand{\dwave}{\ensuremath{D}-wave\xspace}

\renewcommand{\eqref}[1]{Eq.~(\ref{#1})}
\newcommand{\figref}[1]{Fig.~\ref{#1}}
\newcommand{\tabref}[1]{Table~\ref{#1}}
\newcommand{\appref}[1]{Appendix~\ref{#1}}

\newcommand{\secref}[1]{Sec.~\ref{#1}}

\newcommand{\phiM}{\ensuremath{\phi_{\rm M}}\xspace}
\newcommand{\phiD}{\ensuremath{\phi_{\rm D}}\xspace}
\newcommand{\qtag}{\ensuremath{\mathfrak{q}}\xspace}
\newcommand{\qtagT}{\ensuremath{\mathfrak{q}^{\rm X}}\xspace}
\newcommand{\qtagOS}{\ensuremath{\mathfrak{q}^{\rm OS}}\xspace}
\newcommand{\qtagSS}{\ensuremath{\mathfrak{q}^{\rm SS}}\xspace}

\begin{document}

\renewcommand{\thefootnote}{\fnsymbol{footnote}}
\setcounter{footnote}{1}

\begin{titlepage}
\pagenumbering{roman}
\vspace*{-1.5cm}
\centerline{\large EUROPEAN ORGANIZATION FOR NUCLEAR RESEARCH (CERN)}
\vspace*{1.5cm}
\noindent
\begin{tabular*}{\linewidth}{lc@{\extracolsep{\fill}}r@{\extracolsep{0pt}}}
\ifthenelse{\boolean{pdflatex}}
{\vspace*{-2.7cm}\mbox{\!\!\!\includegraphics[width=.14\textwidth]{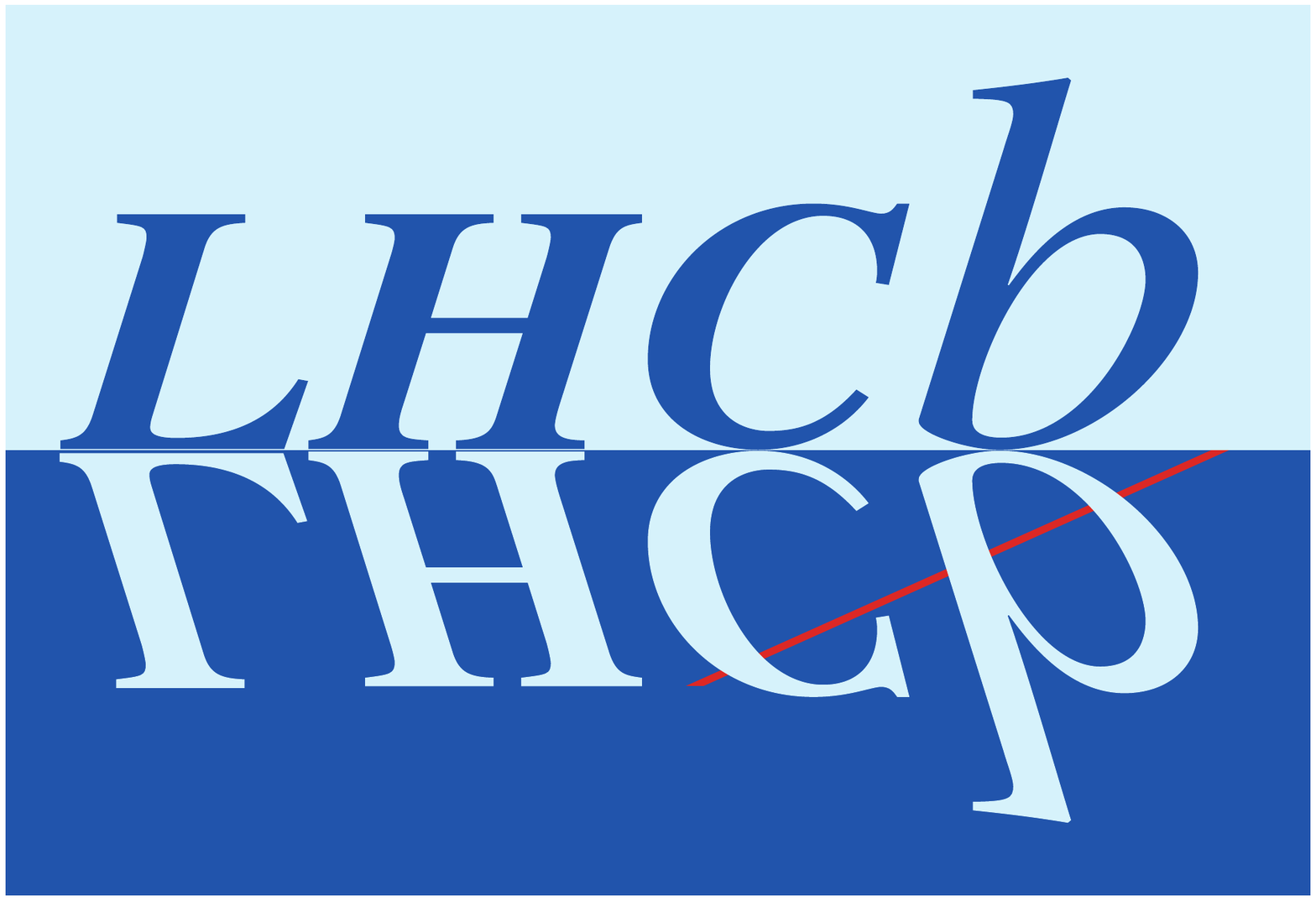}} & &}%
{\vspace*{-1.2cm}\mbox{\!\!\!\includegraphics[width=.12\textwidth]{lhcb-logo.eps}} & &}%
\\
 & & CERN-EP-2017-320 \\
 & & LHCb-PAPER-2017-048 \\
 & & 22 December 2017 \\
\end{tabular*}
\vspace*{4.0cm}
{\normalfont\bfseries\boldmath\huge
\begin{center}
  First measurement of the \CP-violating phase \phisdd in \BsKpiKpi decays
\end{center}
}
\vspace*{1.0cm}
\begin{center}
The LHCb collaboration\footnote{Authors are listed at the end of this paper.}
\end{center}
\vspace*{1.0cm}
\begin{abstract}
\noindent
A flavour-tagged decay-time-dependent amplitude analysis of \BsKpiKpi decays is presented in the \Kpipm mass
range from 750 to 1600\mevcc. The analysis uses $pp$ collision data collected with the LHCb detector at centre-of-mass energies of $7$ and $8\tev$,
corresponding to an integrated luminosity of $3.0\invfb$.
Several quasi-two-body decay modes are considered, corresponding to \Kpm{}\pimp combinations with
spin 0, 1 and 2, which are dominated by the \Kstarzss and \Kstarzs, the \Kstarzp and the \Kstarzd resonances, respectively. The longitudinal polarisation fraction
for the $\Bs\to\Kstarzp\Kstarzbp$ decay is measured as $\fL=0.208 \pm 0.032 \pm 0.046$, where the
first uncertainty is statistical and the second is systematic.
The first measurement of the mixing-induced \CP-violating phase, $\phi_s^{d\bar{d}}$, in $\bquark \to \dquark\dquarkbar\squark$ transitions is performed,
yielding
a value of
  $\phi_s^{d\bar{d}}=-0.10\pm 0.13 \stat \pm 0.14 \syst\rad$.
\end{abstract}
\vspace*{1.0cm}
\begin{center}
  Published in JHEP 03 (2018) 140
\end{center}
\vspace{\fill}
{\footnotesize
\centerline{\copyright~CERN on behalf of the \lhcb collaboration, licence \href{http://creativecommons.org/licenses/by/4.0/}{CC-BY-4.0}.}}
\vspace*{2mm}
\end{titlepage}
\newpage
\setcounter{page}{2}
\mbox{~}
\cleardoublepage

\renewcommand{\thefootnote}{\arabic{footnote}}
\setcounter{footnote}{0}

\pagestyle{plain}
\setcounter{page}{1}
\pagenumbering{arabic}

\section{Introduction}
\label{sec:Introduction}
\begin{figure}[b]
\center
\includegraphics[width=0.6\textwidth]{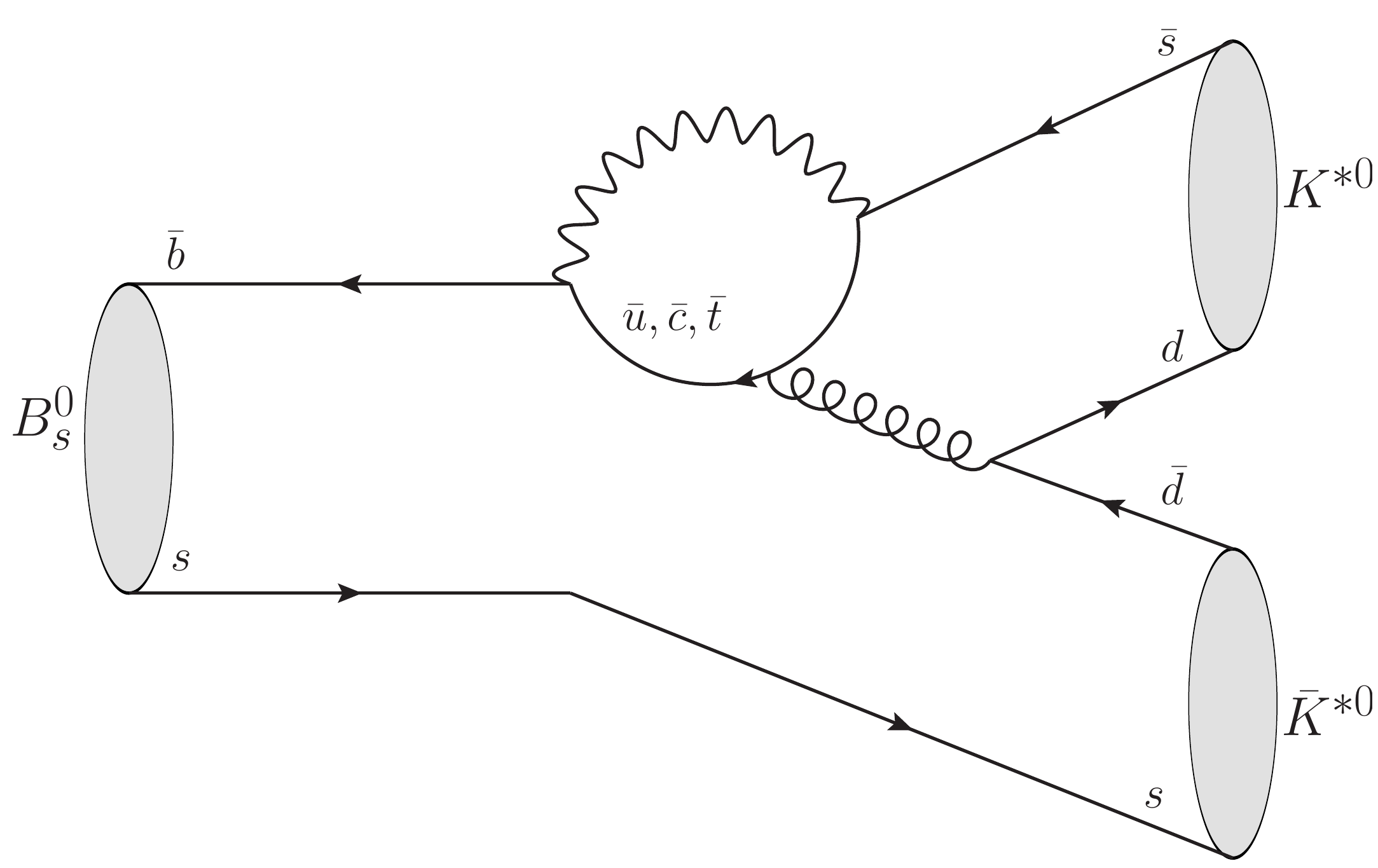}
\caption{Leading-order SM Feynman diagram of the \BsKstKst decay.}
\label{fig:decay_diagram}
\end{figure}

The \CP-violating weak phases \phis arise in the interference between the amplitudes of \Bs mesons
directly decaying to \CP eigenstates
and those decaying to the same final state after \Bs--\Bsb oscillation.  The \BsKstKst
decay,\footnote{Throughout this article, charge conjugation is implied and \Kstarz refers to the \Kstarzp resonance, unless otherwise stated.} which in the Standard Model (SM) is dominated by the
gluonic loop diagram shown in Fig.~\ref{fig:decay_diagram}, has been discussed extensively in the
literature as a benchmark test for the SM and as an excellent probe for physics beyond the
SM~\cite{Fleischer:1999zi, Fleischer:2007wg, Ciuchini:2007hx, DescotesGenon:2007qd,
DescotesGenon:2011pb, Bhattacharya:2012hh, Bhattacharya:2013sga}.
New heavy particles entering the loop would
introduce additional amplitudes and modify properties of the decay from their SM values.
In general, the weak phase \phis depends on the \Bs decay channel under consideration, and can be
different between channels as it depends on the contributions from tree- and loop-level processes.  The notation
\phisdd is used when referring to the weak phase measured in \btoddbs transitions. For \btoccbs
transitions, {\em e.g.}~$\Bs \to \jpsi \Kp \Km$ and $\Bs \to \jpsi \pip \pim$ decays, the weak
phase \phiscc has been measured by several experiments~\cite{HFLAV16, LHCb-PAPER-2014-059,
Aad:2016tdj, Khachatryan:2015nza}. The world average reported by HFLAV, $\phiscc = -0.021 \pm
0.031\rad$~\cite{HFLAV16}, is dominated by the LHCb measurement $\phiscc = -0.010 \pm 0.039\rad$~\cite{LHCb-PAPER-2014-059}.
The LHCb collaboration has also measured the \phisss phase in $\Bs
\to \phi\phi$ transitions~\cite{LHCb-PAPER-2014-026}, reporting a value of $\phisss = -0.17 \pm
0.15\rad$.
The decay \BsKstKst, with $\Kstarz\to K^+\pi^-$ and $\Kstarzb\to K^-\pi^+$, was first observed by
the LHCb collaboration, based on $pp$
collision data corresponding to an integrated
luminosity of $35\invpb$ at a centre-of-mass energy $\sqrt{s} = 7\tev$~\cite{LHCb-PAPER-2011-012}. A branching fraction and a final-state polarisation analysis were reported.
An updated analysis of
the \BsKpiKpi decay was performed by LHCb using $1.0$\invfb
of data at $\sqrt{s} = 7\tev$~\cite{LHCb-PAPER-2014-068}. In both analyses, the invariant mass of the two $K\pi$ pairs\footnote{Hereafter the
notation $K\pi$ will stand for both $K^+\pi^-$ and $K^-\pi^+$ pairs.} was restricted to a window of $\pm
150$\mevcc around the known \Kstarz mass.
This publication reports the first decay-time-dependent amplitude analysis of \BsKpiKpi decays using a $K\pi$ mass window that extends from 750
to 1600\mevcc, approximately corresponding to the region between the $K\pi$
production threshold and the $\Dz\to \Km\pip$ resonance. At the current level of sensitivity, the assumption of common \CP-violating parameters for the contributing amplitudes is appropriate.
Consequently, such a wide window provides a four-fold increase of the signal sample size with
respect to the narrow window of $150\mevcc$ around the \Kstarz mass.
The analysis uses $pp$ collision data collected by LHCb in 2011 and 2012 at $\sqrt{s}=7$ and $8\tev$,
corresponding to an integrated luminosity of $3.0\invfb$.
In this study, nine different quasi-two-body decay channels are considered, corresponding to the
different possible combinations of $K\pi$ pairs with spin 0, 1 or 2. Additional contributions were
studied and found to be negligible in the phase-space region considered in this analysis.  The
$K\pi$ spectrum is dominated by the $K_0^*(800)^0$, \Kstarzs, \Kstarzp and \Kstarzd resonances. Angular
momentum conservation in the decay allows for one single amplitude in modes involving at least one
scalar $K\pi$ pair, three amplitudes for vector-vector or vector-tensor decays and five amplitudes
for a tensor-tensor decay. These possibilities are listed in Table~\ref{tab:channels}. There is a physical difference between decay pairs of the form scalar-vector and vector-scalar. Namely, in the used convention, the spectator quark from the $B_s^0$ decay (see \figref{fig:decay_diagram}) always ends up in the second $K\pi$ pair.
The \CP-averaged fractions of the contributing amplitudes, $f_i$, as well as their strong-phase
differences, $\delta_i$, are determined together with the \CP-violating weak phase \phisdd and a
parameter that accounts for the amount of \CP violation in decay, $|\lambda|$. This is the first
time that the weak phase in \btoddbs transitions has been measured. It is also the first
time that the tensor components in the $(K^+\pi^-)(K^-\pi^+)$ system have been studied.
\begin{table}[t]
\centering
\caption{Quasi-two-body decay channels and corresponding polarisation amplitudes contributing to the \BsKpiKpi final state in the $K\pi$ mass window from 750 to 1600\mevcc. The different contributions are identified by the spin $j_1$ ($j_2$) of the $K^+\pi^-$ ($K^-\pi^+$) pair and the helicity $h$. In cases where more than one amplitude contributes, the polarisations are defined as being longitudinal,
parallel, or perpendicular, which are then denoted by $0$, $\parallel$ and $\perp$ respectively, following the definitions given in
  Ref.~\cite{PDG2016}. The subscripts 1 and 2 in the parallel and perpendicular helicities of the tensor-tensor component denote different spin states leading
  to a parallel or a
  perpendicular configuration, as discussed in \appref{sec:angular}.}
  \renewcommand{\arraystretch}{1.3}
  \begin{tabular}{l l c c x{3cm} x{3cm} }
   Decay & Mode & $j_1$ & $j_2$ & Allowed values of $h$ & Number of amplitudes \\
   \hline
   $\Bs \to (K^{+}\pi^{-})^{*}_{0}(K^{-}\pi^{+})^{*}_{0}$ & scalar-scalar  & 0 & 0 & 0                                                       & 1 \\
   $\Bs \to (K^{+}\pi^{-})^{*}_{0}\Kstarzbp$              & scalar-vector  & 0 & 1 & 0                                                       & 1 \\
   $\Bs \to \Kstarzp (K^{-}\pi^{+})^{*}_{0}$              & vector-scalar  & 1 & 0 & 0                                                       & 1 \\
   $\Bs \to (K^{+}\pi^{-})^{*}_{0}\Kstarzbd$              & scalar-tensor  & 0 & 2 & 0                                                       & 1 \\
   $\Bs \to \Kstarzd (K^{-}\pi^{+})^{*}_{0}$              & tensor-scalar  & 2 & 0 & 0                                                       & 1 \\
   $\Bs \to \Kstarzp \Kstarzbp$                           & vector-vector  & 1 & 1 & 0, $\parallel$, $\perp$                                 & 3 \\
   $\Bs \to \Kstarzp \Kstarzbd$                           & vector-tensor  & 1 & 2 & 0, $\parallel$, $\perp$                                 & 3 \\
   $\Bs \to \Kstarzd \Kstarzbp$                           & tensor-vector  & 2 & 1 & 0, $\parallel$, $\perp$                                 & 3 \\
   $\Bs \to \Kstarzd \Kstarzbd$                           & tensor-tensor  & 2 & 2 & 0, $\parallel_1$, $\perp_1$, $\parallel_2$, $\perp_2$   & 5 \\
\end{tabular}
 \label{tab:channels}
\end{table}
\section{Phenomenology}
\label{sec:Phenomenology}
\begin{figure}[t]
\center
\includegraphics[width=100mm]{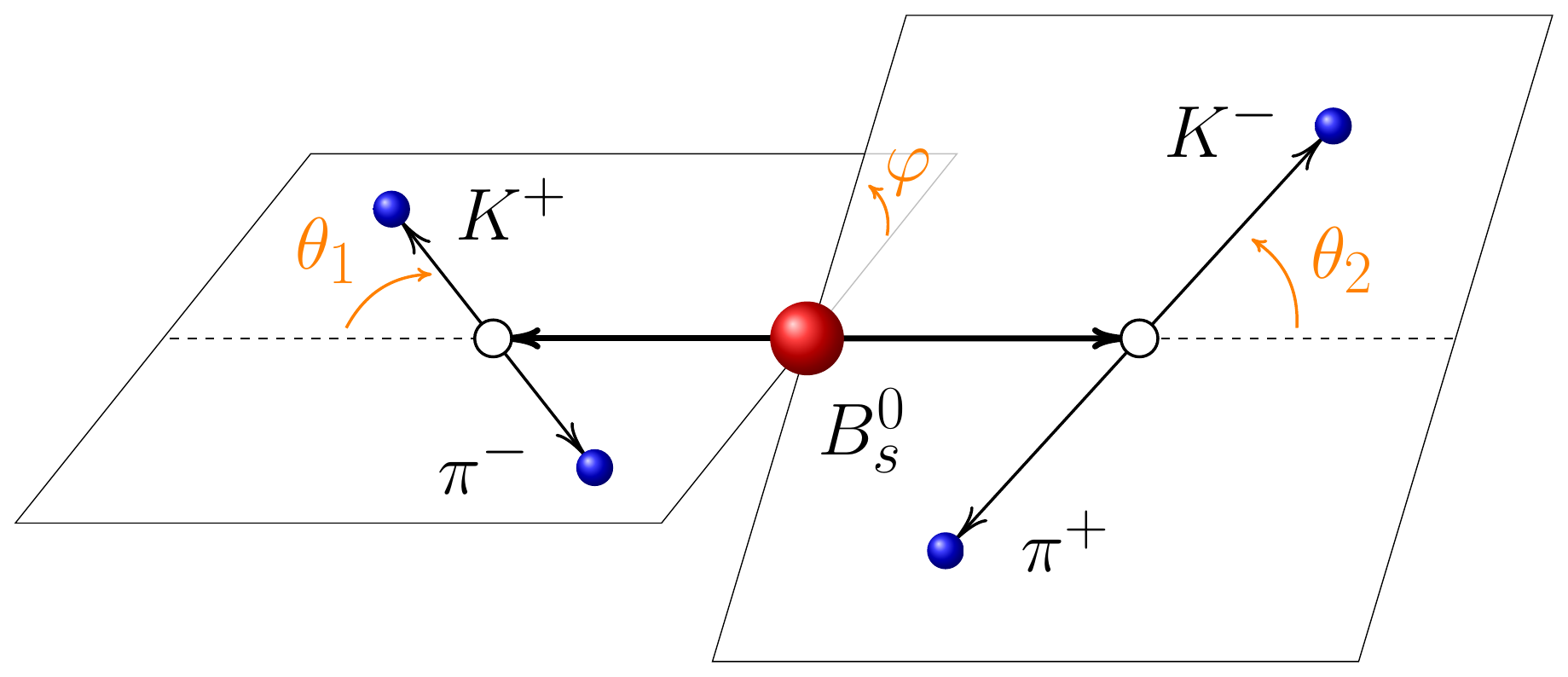}
\caption{Graphical definition of the angles in the helicity basis. Taking the example of a $\Bs \to Q_1
Q_2$ decay (this analysis uses $\Bs\to SS$, $\Bs\to SV$, $\Bs\to VS$, $\Bs\to VV$,
$\Bs\to ST$, $\Bs\to TS$, $\Bs\to VT$, $\Bs\to TV$ and $\Bs\to TT$), with each final-state
quasi-two-body meson decaying to pseudoscalars ($Q_1 \to K^+\pi^-$ and $Q_2 \to
K^-\pi^+$), $\theta_1$ ($\theta_2$) is defined as the angle between the directions of
motion of $K^+$ ($K^-$) in the $Q_1$ ($Q_2$) rest frame and $Q_1$ ($Q_2$) in the \Bs rest frame,
and $\varphi$ as the angle between the plane defined by $K^+\pi^-$ and the plane defined by
$K^-\pi^+$ in the \Bs rest frame.}
\label{fig:anghelbasis}
\end{figure}

The phenomenon of quark mixing means that a \Bs meson can oscillate into its antiparticle equivalent,
\Bsb. Consequently, the physical states, \BsH (heavy) and \BsL (light), which have mass and decay
width differences defined by $\Deltams = \mBH - \mBL$ and $\DeltaGammas = \GammaBL - \GammaBH$,
respectively, are admixtures of the flavour eigenstates such that
\begin{equation}
  \BsH = p\Bs + q\Bsb \;\;\;\;\; \mathrm{and} \;\;\;\;\; \BsL = p\Bs - q\Bsb,
  \label{eq:mass_states}
\end{equation}
where $p$ and $q$ are complex coefficients that satisfy $|p|^2 + |q|^2 = 1$.
The time evolution of the initially pure flavour eigenstates at $t=0$, $\ket{\Bs(0)}$ and $\ket{\Bsb(0)}$, is described by
\begin{equation}
\begin{aligned}
\ket{\Bs(t)}&=g_+(t)\ket{\Bs(0)}+\frac{q}{p}g_-(t)\ket{\Bsb(0)},\\
\ket{\Bsb(t)}&=\frac{p}{q}g_-(t)\ket{\Bs(0)}+g_+(t)\ket{\Bsb(0)},
\end{aligned}
\end{equation}
where the decay-time-dependent functions $g_{\pm}(t)$ are given by
\begin{equation}
g_{\pm}(t)=\frac{1}{2}e^{-i\ms t}e^{-\frac{\Gammas}{2}t}\left(e^{i\frac{\Deltams}{2}t}e^{-\frac{\DeltaGammas}{4}t}\pm e^{-i\frac{\Deltams}{2}t}e^{\frac{\DeltaGammas}{4}t}\right),
\end{equation}
with \ms and \Gammas being the average mass and width of the \BsH and \BsL states. Negligible
\CP violation in mixing is assumed in this analysis, leading to the parameterisation
$q/p=e^{-i\phiM}$, where $\phiM$ is the \Bs--\Bsb mixing phase.
The total decay amplitude of the flavour eigenstates at $t=0$ into the final state $f=(\Kp\pim)(\Km\pip)$, denoted by
$\braket{f}{\Bs(0)}$ and $\braket{f}{\Bsb(0)}$, is a coherent sum of scalar-scalar (SS), scalar-vector (SV), vector-scalar (VS), scalar-tensor (ST), tensor-scalar (TS),
vector-vector (VV), vector-tensor (VT), tensor-vector (TV) and tensor-tensor (TT) contributions.
The quantum numbers used to label the $(\Kp\pim)(\Km\pip)$ final states
are the spin $j_1$ ($j_2$) of the $K^+\pi^-$ ($K^-\pi^+$) pair and the helicity $h$.
The vector component is represented in this analysis by the $K^{*0}$
meson, since this resonance is found to be largely dominant in this spin configuration.
Potential contributions from the $K_1^*(1410)^0$ and $K_1^*(1680)^0$ resonances are considered as
sources of systematic uncertainty. For the tensor case, only the $K_2^*(1430)^0$ resonance contributes in
the considered $K\pi$ mass window. The scalar component, denoted in this paper by $(K\pi)_0^*$
requires a more careful treatment. It can have contributions from the $K_0^*(800)^0$ and
$K_0^*(1430)^0$ resonances and from a nonresonant $K\pi$ component. The parameterisation of the
$K\pi$ invariant mass spectrum for the scalar contribution is explained later in this section. All
of the considered decay modes, together with the quantum numbers for the corresponding amplitudes,
are shown in \tabref{tab:channels}.
In order to separate components with different \CP eigenvalues, $\eta_h^{j_1j_2}=\pm1$, the
differential decay rate is expressed as a function of three angles and the two $K\pi$
invariant masses. The angles $\theta_1$, $\theta_2$ and $\varphi$, are written in the
helicity basis and defined according to the diagram shown in \figref{fig:anghelbasis}. The
invariant mass of the \Kp\pim pair is denoted as $m_1$, while that of the \Km\pip pair as $m_2$.
The symbol $\Omega$ is used to represent all three angles and the two invariant masses,
$\Omega = (m_1,m_2,\cos\theta_1,\cos\theta_2,\varphi)$.
Summing over the possible states and using the partial wave formalism, the decay amplitudes at $t=0$ can be written as
\begin{equation}
\begin{aligned}
\label{eq:angmassamplitudes}
\braket{f}{\Bs(0)}(\Omega)&=\sum_{j_1,j_2,h}\mathcal{A}_{h}^{j_1 j_2}\Theta^{j_1 j_2}_{h}(\cos{\theta_1},\cos{\theta_2},\varphi)\mathcal{H}_{h}^{j_1 j_2}(m_1,m_2),\\
\braket{f}{\Bsb(0)}(\Omega)&=\sum_{j_1,j_2,h}\eta_h^{j_1 j_2}\overline{\mathcal{A}}_{h}^{j_1 j_2}\Theta^{j_1 j_2}_{h}(\cos{\theta_1},\cos{\theta_2},\varphi)\mathcal{H}_{h}^{j_1 j_2}(m_1,m_2).
\end{aligned}
\end{equation}
The complex parameters $\mathcal{A}_{h}^{j_1 j_2}$ and $\overline{\mathcal{A}}_{h}^{j_1 j_2}$ contain the physics of the decays to the final states with $j_1$, $j_2$ and $h$ as
defined in Table~\ref{tab:channels}.  The angular terms, $\Theta_h^{j_1 j_2}$, are built from
combinations of spherical harmonics as shown in \appref{sec:angular}. The $\eta_h^{j_1 j_2}$ factor
is equal to $(-1)^{j_1+j_2}\,\eta_h$, where $\eta_h=1$ for
$h\in\{0,\parallel,\parallel_1,\parallel_2\}$ and $\eta_h=-1$ for $h\in\{\perp,\perp_1,\perp_2\}$.
The mass-dependent terms are parameterised as
\begin{equation}
  \label{eq:massdependamps}
\mathcal{H}_{h}^{j_1 j_2}(m_1,m_2)=\mathcal{F}_h^{j_1 j_2}(m_1,m_2)\mathcal{M}_{j_1}(m_1)\mathcal{M}_{j_2}(m_2),
\end{equation}
where $\mathcal{F}_h^{j_1 j_2}(m_1,m_2)$ is the Blatt--Weisskopf angular-momentum
centrifugal-barrier factor~\cite{VonHippel:1972fg} and $\mathcal{M}_{j}$ describes the shape of the
$K\pi$ invariant mass of a $K\pi$ pair with spin $j$. Relativistic Breit--Wigner
functions of spin 1 and 2, parameterising the \Kstarz and the \Kstarzd
resonances, are used for $\mathcal{M}_1$ and $\mathcal{M}_2$, respectively.
The parameterisation of $\mathcal{M}_0$ is based on the phenomenological \swave scattering
amplitude of isospin $1/2$ presented in Ref.~\cite{PhysRevD.93.074025}. Since only the phase evolution
of $\mathcal{M}_0$ is linked to that of the scattering amplitude (by virtue of Watson's
theorem~\cite{PhysRev.95.228}), its modulus is parameterised with a fourth-order
polynomial whose coefficients are determined in the final fit to data. Details of this
parameterisation can be found in \appref{sec:M0}.
The normalisation condition for the mass-dependent terms is
\begin{equation}
\int d m_1 \int d m_2\;|\mathcal{H}_{h}^{j_1 j_2}(m_1,m_2)|^2\Phi_4(m_1,m_2)=1,
\label{eq:norm}
\end{equation}
where $\Phi_4$ is the four-body phase-space factor. The phase of $\mathcal{H}_{h}^{j_1
j_2}(m_1,m_2)$ is set to 0 at $m_1=m_2=M(\Kstarz)$, where $M(\Kstarz)$ is the mass of the
$\Kstarz$ state~\cite{PDG2016}, in order to normalise the relative global phases of the $K\pi$ mass-dependent amplitudes.
The \CP-violating effects are assumed to be the same for all of the modes under study.
Consequently, the value of \phisdd and $|\lambda|$ determined in this
article is effectively an average over the various channels considered in Table~\ref{tab:channels}.
Within this approach, the physical amplitudes $\mathcal{A}_{h}^{j_1 j_2}$ and
$\overline{\mathcal{A}}_{h}^{j_1 j_2}$ in \eqref{eq:angmassamplitudes} can be separated into a
\CP-averaged complex amplitude, $A_h^{j_1 j_2}$,
a direct \CP asymmetry, $\Delta^{\CP}_{\rm dir}=(|\overline{\mathcal{A}_h^{j_1 j_2}}|^2 -
|\mathcal{A}_h^{j_1 j_2}|^2)/(|\overline{\mathcal{A}_h^{j_1 j_2}}|^2 + |\mathcal{A}_h^{j_1
j_2}|^2)$,\footnote{The direct \CP asymmetry is often notated elsewhere as $A_{\CP}$.} and a \CP-violating weak phase in the decay, $\phiD$, as
\begin{equation}
\label{eq:CPVandCPAv}
\begin{aligned}
\mathcal{A}_h^{j_1 j_2}&=\sqrt{1-\Delta^{\CP}_{\rm dir}} e^{-i\phiD} A_h^{j_1 j_2},\\
\overline{\mathcal{A}}_h^{j_2 j_1}=\overline{\mathcal{A}_h^{j_1 j_2}}&=\sqrt{1+\Delta^{\CP}_{\rm dir}} e^{i\phiD} A_h^{j_1 j_2}.
\end{aligned}
\end{equation}
In the expressions above the \CP transformation also changes $j_1j_2$ to
$j_2j_1$. The total \CP-violating phase associated to the interference between mixing and decay is
given by $\phisdd =\phiM-2\phiD$ and its determination is the main goal of this analysis. In the SM the size of \phisdd is expected to be small due to an almost exact
cancellation in the values of $\phiM$ and $2\phiD$~\cite{DescotesGenon:2011pb}.
The parameter $|\lambda|$ is defined in terms of the direct \CP asymmetry by
\begin{equation}
|\lambda|=\frac{\sqrt{1+\Delta^{\CP}_{\rm dir}}}{\sqrt{1-\Delta^{\CP}_{\rm dir}}}.
\end{equation}
\section{Detector and simulation}
\label{sec:Detector}
The \lhcb detector~\cite{Alves:2008zz,LHCb-DP-2014-002} is a single-arm forward spectrometer covering the
\mbox{pseudorapidity} range between $2$ and $5$, designed for the study of particles containing \bquark or \cquark quarks. The
detector includes a high-precision tracking system consisting of a silicon-strip vertex detector surrounding the $pp$
interaction region, a large-area silicon-strip detector located upstream of a dipole
magnet with a bending power of about $4{\mathrm{\,Tm}}$, and three stations of silicon-strip detectors and straw drift
tubes placed downstream of the magnet.  The tracking system provides a measurement of
momentum, \ptot, of charged particles with relative uncertainty that varies from 0.5\% at low momentum to 1.0\% at
200\gevc.  The minimum distance of a track to a primary vertex (PV), the impact parameter (IP), is measured with
resolution of $(15+29/\pt)\mum$, where \pt is the component of the momentum transverse to the beam, in\,\gevc.
Different types of charged hadrons are distinguished using information from two ring-imaging Cherenkov (RICH)
detectors.  Photons, electrons and hadrons are identified by a calorimeter system
consisting of scintillating-pad and preshower detectors, an electromagnetic calorimeter and a hadronic calorimeter.
Muons are identified by a system composed of alternating layers of iron and multiwire proportional
chambers.
The online event selection is performed by a
trigger, which consists of a hardware stage, based on information from the calorimeter
and muon systems, followed by a software stage, which applies a full event reconstruction.
At the hardware trigger stage, events are required to contain a muon with high \pt or a
hadron, photon or electron with high transverse energy in the calorimeters.
The software trigger requires a two-, three- or four-track
secondary vertex with significant displacement from the primary
$pp$ interaction vertices. At least one charged particle
must have transverse momentum $\pt > 1.7\gevc$ and be
inconsistent with originating from a PV.
A multivariate algorithm~\cite{BBDT} is used for
the identification of secondary vertices consistent with the decay
of a \bquark hadron.
Simulated samples of resonant
{\BsKstKst}, {\BsKstKsts} and {\BsKstsKsts} decays, as well as phase-space {\BsKpiKpiPS} decays,
are used to study the signal. Simulated samples of {\BdKstKst}, {\BdKstRho}, {\BdKstPhi} and {\LbPKpipi}
are created to study peaking backgrounds.
In the simulation, $pp$ collisions are generated using
\pythia~\cite{Sjostrand:2007gs}
 with a specific \lhcb
configuration~\cite{LHCb-PROC-2010-056}.  Decays of particles
are described by \evtgen~\cite{Lange:2001uf}, in which final-state
radiation is generated using \photos~\cite{Golonka:2005pn}. The
interaction of the generated particles with the detector, and its response,
are implemented using the \geant
toolkit~\cite{Allison:2006ve, *Agostinelli:2002hh} as described in
Ref.~\cite{LHCb-PROC-2011-006}.
\section{Signal candidate selection}
\label{sec:Selection}
\begin{figure}
  \includegraphics[width=0.49\textwidth]{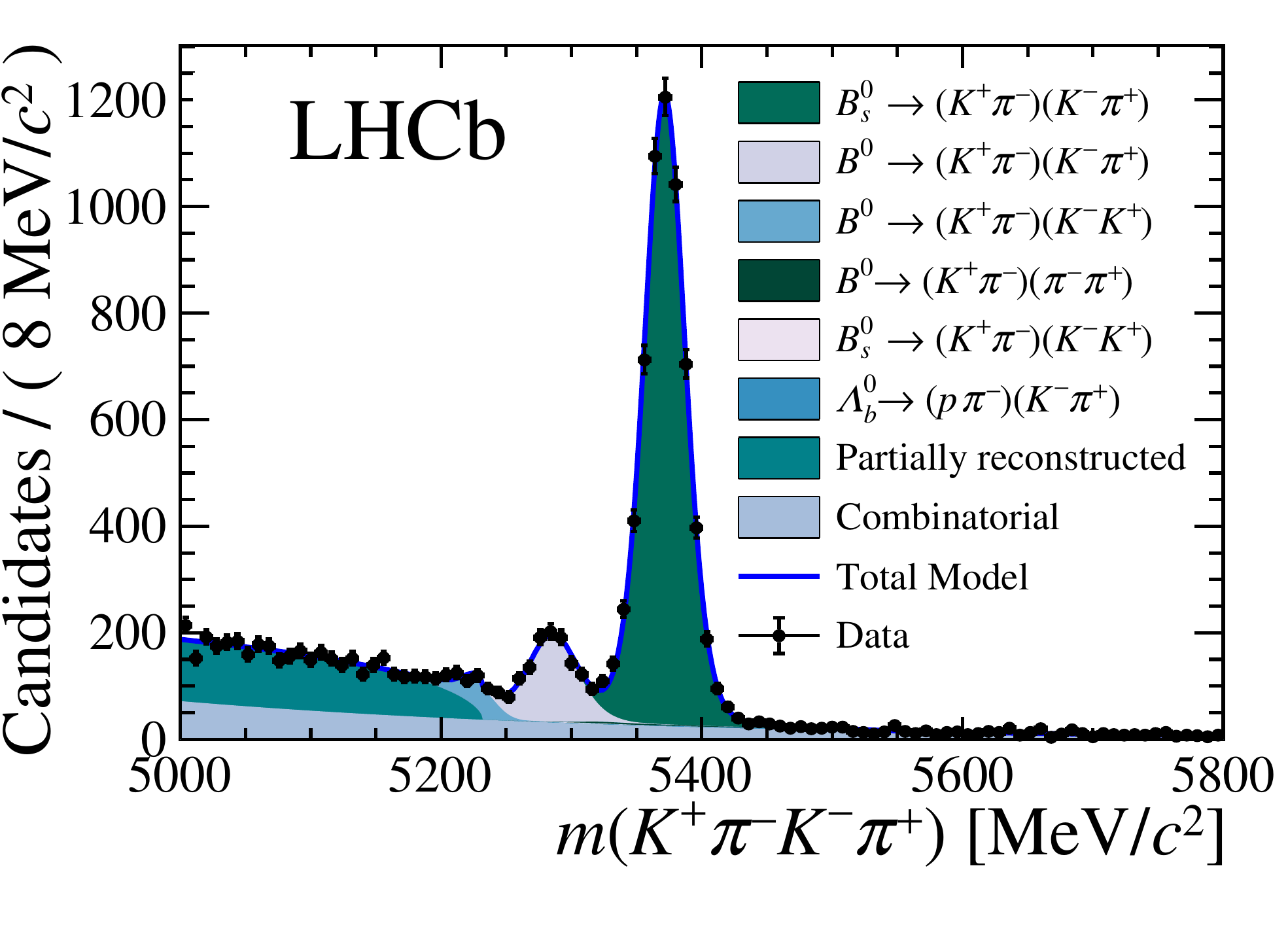}
  \includegraphics[width=0.49\textwidth]{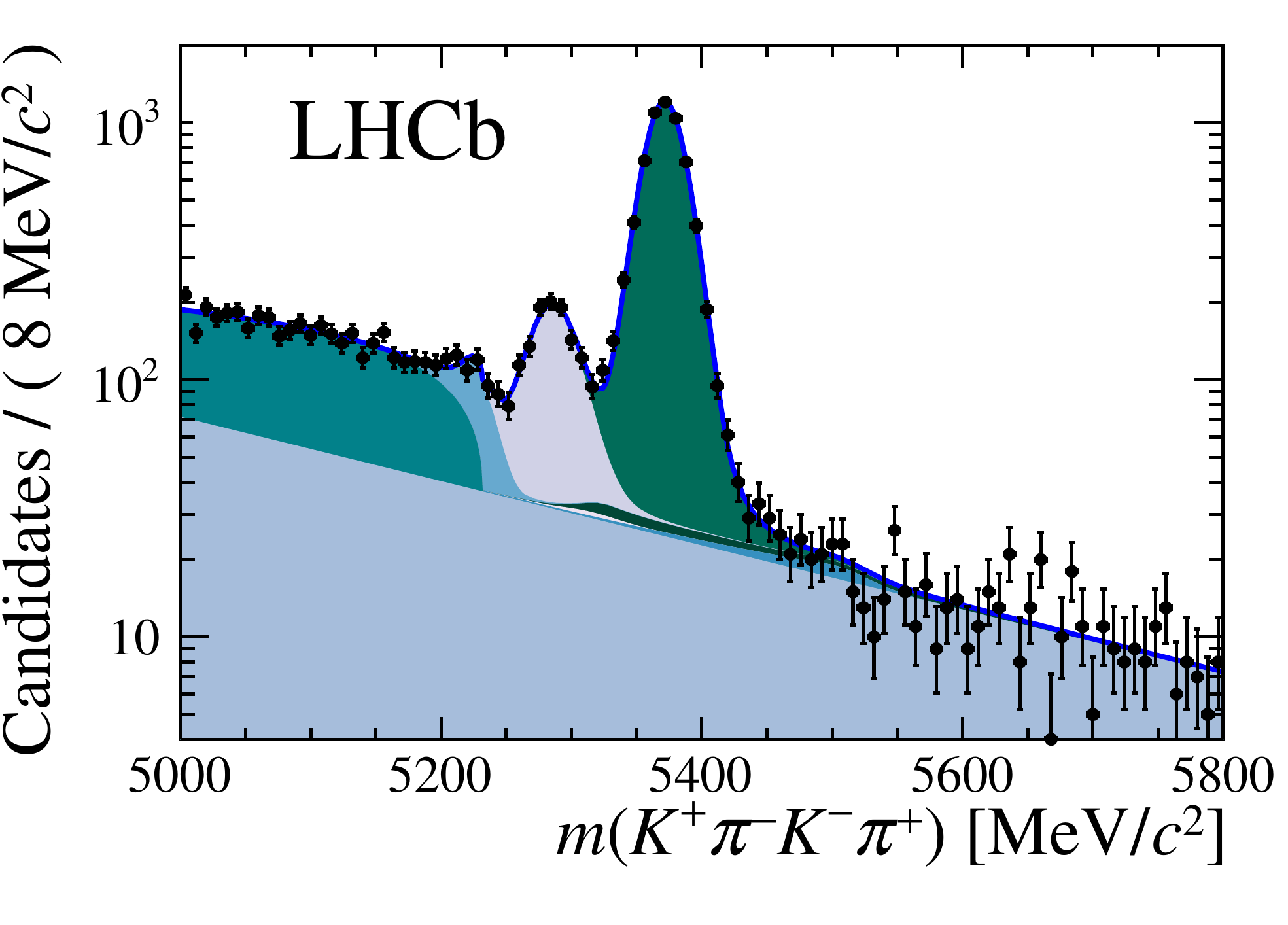}
  \caption{Four-body invariant mass distribution on a (left) linear and (right) logarithmic scale superimposed with the mass fit model.}
  \label{fig:inv_mass_fit}
\end{figure}

\begin{figure}
  \centering
  \includegraphics[width=0.7\textwidth]{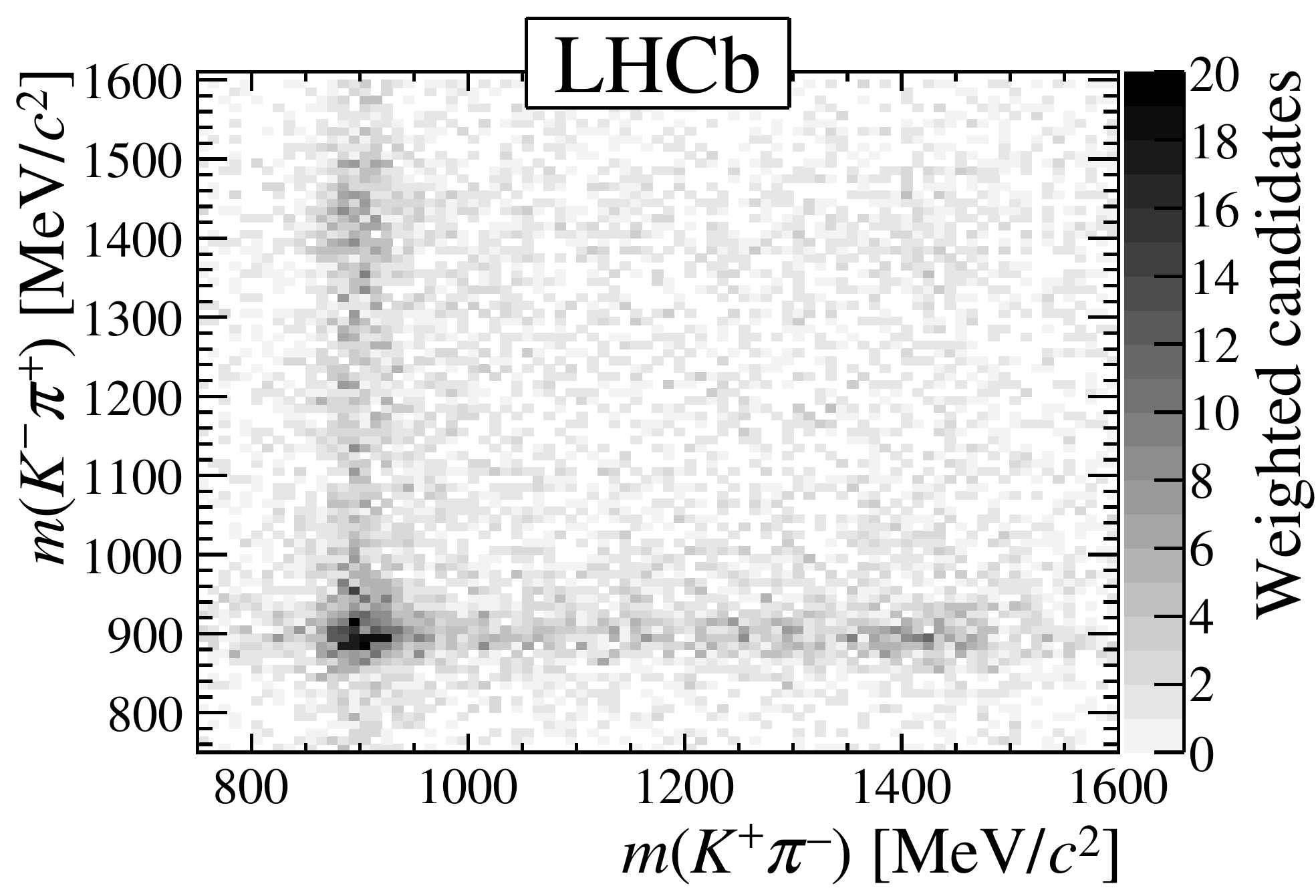}
  \caption{Distribution of the two $(K\pi)$ pair invariant masses, with the signal weights applied, after all of the selection requirements.}
  \label{fig:dalitz}
\end{figure}

Events passing the trigger are required to satisfy requirements on the fit quality of the \Bs decay vertex as well as the \pt
and \chisqip of each
track, where \chisqip is defined as the difference between the \chisq of the secondary vertex reconstructed with and without the track under consideration.
The tracks are assigned as kaon or pion candidates using
particle identification information from the RICH detectors by requiring that the likelihood for the kaon hypothesis is larger
than that for the pion hypothesis and vice versa.
In addition, the \pt of each $K\pi$ pair is required to be larger than $500$\mevc, the reconstructed mass of
each $K\pi$ pair is required to be within the range $750\leq m(K\pi) \leq 1600$\mevcc and the reconstructed mass of the
\Bs candidate
is required to be within the range $5000\leq m(\Kp\pim\Km\pip) \leq 5800$\mevcc.
A boosted decision tree (BDT) algorithm~\cite{Breiman,AdaBoost} is trained to reject combinatorial background, where at least one of the final-state tracks
originates from a different decay or directly from the PV. The signal is represented in the BDT training with simulated
$\Bs\to\Kstarz\Kstarzb$ candidates, satisfying the same requirements as the data, while selected data candidates in the four-body invariant
mass sideband, $5600 \leq m(\Kp\pim\Km\pip) \leq 5800$\mevcc, are used to represent the background. The input variables employed in the training are
kinematic and geometric quantities associated with the four final-state tracks, the two $K\pi$ candidates and the \Bs candidate.
The features used to train the BDT response are chosen to minimise any correlation with the \Bs and two $K\pi$ pair invariant masses.
Separate trainings are performed for the data samples collected in 2011 and 2012, due
to the different data-taking conditions. The $k$-fold cross-validation method~\cite{Blum:1999bhb}, with $k=4$, is used to
increase the training statistics while reducing the risk of overtraining. The requirement on the BDT response is optimized by maximising the metric
$N_{\rm S}/\sqrt{N_{\rm S}+N_{\rm B}}$, where $N_{\rm S}$ is the estimated number of signal candidates after selection
and $N_{\rm B}$ is the estimated number of combinatorial background
candidates within $\pm 60\mevcc$ of the known \Bs mass~\cite{PDG2016}.
The BDT requirement is 95\% efficient for simulated signal candidates and rejects 70\% of the combinatorial background.
After applying the BDT requirement, specific background contributions containing two real oppositely charged kaons and two real oppositely charged pions are removed by mass vetoes on the two- and
three-body invariant masses. Candidates are removed if they fulfill either $m(\Kp\Km\pipm)<2100$\mevcc or $m(\Kp\Km)$ within 30\mevcc of the
known \Dz mass~\cite{PDG2016}.
Sources of peaking background in which one of the final-state tracks is misidentified are
suppressed by introducing further particle identification requirements.
The particle identification quantities make use of information from the RICH detectors and are calibrated using
$\Dstarp\to\Dz\pip$ and $\Lb\to\Lc\mun\neumb$ decays in data.
These requirements significantly reduce contributions from $\Bd\to\rhoz\Kstarz$, $\Bd\to\phi\Kstarz$ and
$\Lb\to\proton\pim\Km\pip$ in which a pion or proton is misidentified as a kaon, or a kaon is misidentified as a pion. In addition, there are
specific extra particle identification requirements for candidates whose reconstructed mass falls within $\pm30$\mevcc of the known \Bd or \Lb mass under
the relevant mass-hypothesis change ($K\to\proton$, $K\to\pi$ or $\pi\to K$). These requirements remove 40\% of the simulated signal
but almost all of the simulated background:
80\% of $\Bd\to\phi\Kstarz$, 96\% of $\Bd\to\rho\Kstarz$ and 88\% of $\Lb\to\proton\pim\Km\pip$ events.
Subsequently, each of these background components is found to have a small effect on the signal determination.
After all of the selection criteria have been imposed, 1.4\% of selected events contain multiple candidates, from which one is randomly selected.
A fit to the four-body invariant mass distribution is performed in order to determine a set of signal weights, obtained using
the \sPlot procedure~\cite{Pivk:2004ty}, which allows the decay-time-dependent \CP fit to be performed on a sample that represents only the signal. For the
invariant mass fit the $\Bs\to(\Kp\pim)(\Km\pip)$ signal and the peaking background components $\Bd\to(\Kp\pim)(\Km\pip)$,
$B^{0}_{(s)}\to(\Kp\Km)(\Km\pip)$, $\Bd\to(\pip\pim)(\Km\pip)$ and $\Lb\to(\proton\pim)(\Km\pip)$ are modelled as Ipatia functions~\cite{Santos:2013gra}
in which the tail parameters are fixed to values obtained from fits to the simulated samples.
The mass difference between the \Bs and \Bd mesons is fixed to its known value whilst the mean of the \Bs component, as well as
the width of both the \Bd and \Bs components,
are
allowed to vary freely. The yields of the ${\BsKpiKpi}$,
${\Bd\to(\Kp\pim)(\Km\pip)}$ and ${\Bd\to(\Kp\Km)(\Km\pip)}$ components are allowed to freely vary, whilst the yields of the other components are Gaussian
constrained to values relative to the known \mbox{$\Bd\to(\Kp\Km)(\Km\pip)$} branching fraction taking into account the relevant production
fractions~\cite{fsfd} and reconstruction efficiencies. There is an additional background contribution in the low-mass region from partially reconstructed $b$-hadron decays in which
a pion is missed in the final state.
This component is modelled
as an ARGUS function~\cite{ArgusFunction} convolved with a Gaussian mass resolution function. The ARGUS cutoff parameter is fixed to the fitted \Bs mass minus the neutral pion mass, with the other parameters and yield allowed to vary. The
combinatorial background is modelled as an exponential function whose shape parameter and yield are allowed to vary.
The result of the four-body invariant mass fit, which is used to obtain the \sPlot signal weights, is shown in Fig.~\ref{fig:inv_mass_fit}.
The two $K\pi$ pair invariant masses, with the signal weights applied, are shown in Fig.~\ref{fig:dalitz}. The resulting yields of the
various fit components are shown in Table~\ref{tab:fit_yields}.
\begin{table}
  \centering
  \caption{Yields of the signal decay and the various background components considered in the four-body invariant mass fit. The uncertainties are statistical only. The signal region is defined as $\pm 60$\mevcc from the known \Bs meson mass~\cite{PDG2016}.}
  \label{tab:fit_yields}
  \setlength{\tabcolsep}{2pt}
  \renewcommand{\arraystretch}{1.1}
  \begin{tabular}{ p{5cm}  k{1.2cm} c p{1cm} c }
    Channel & \multicolumn{3}{c}{Yield} & Yield in Signal Region \\
    \hline
    $\Bs\to(\Kp\pim)(\Km\pip)$     & 6080 & $\pm$ & 83  & 6004 \\
    $\Bd\to(\Kp\pim)(\Km\pip)$     & 1013 & $\pm$ & 49  & 103  \\
    $\Bd\to(\Kp\pim)(\Km\Kp)$      & 281  & $\pm$ & 47  & 1    \\
    $\Bs\to(\Kp\pim)(\Km\Kp)$      & 8    & $\pm$ & 3   & 4    \\
    $\Bd\to(\Kp\pim)(\pim\pip)$    & 57   & $\pm$ & 13  & 33   \\
    $\Lb\to(\proton\pim)(\Km\pip)$ & 44   & $\pm$ & 10  & 13   \\
    Partially reconstructed        & 2580 & $\pm$ & 151 & 0    \\
    Combinatorial                  & 2810 & $\pm$ & 214 & 372  \\
  \end{tabular}
\end{table}
\section{Flavour tagging}
\label{sec:Tagging}
At the LHC, \bquark quarks are predominantly produced in \bbbar pairs. This analysis focuses on events where one of the quarks hadronises to produce the \Bs meson while the other quark hadronises and decays
independently. Taking advantage of this effect, two types of tagging algorithms aimed at
identifying the \bquark-quark flavour at production time are used in this analysis: same-side (SS)
taggers, based on information from accompanying particles associated with the signal \Bs
hadronisation process; and opposite-side (OS) taggers, based on particles produced in the decay of
the other \bquark quark. This analysis uses the neural-network-based SS-kaon tagging algorithm
presented in Ref.~\cite{LHCb-PAPER-2015-056}; and the combination of OS tagging algorithms explained
in Ref.~\cite{LHCb-PAPER-2011-027}, based on information from \bquark-hadron decays to
electrons, muons or kaons and the total charge of tracks that form a vertex.
Both the SS and OS tagging algorithms provide for each event a tagging decision, \qtag, and an
estimated mistag probability, $\eta_{\rm tag}$. The tagging decision takes the value $1$ for \Bs,
$-1$ for \Bsb and $0$ for untagged. To obtain the calibrated mistag probability for a \Bs(\Bsb)
meson, $\omega$\,($\bar{\omega}$), the estimated probability is calibrated on several
flavour-specific control channels. The following linear functions are used in the calibration
\begin{equation}
\begin{aligned}
\omega^{\rm X} (\eta_{\rm tag}^{\rm X} )&=\left(p_{0}^{\rm X} +\frac{\Delta p_{0}^{\rm X}
}{2}\right)+\left(p_{1}^{\rm X} +\frac{\Delta p_{1}^{\rm X} }{2}\right)(\eta_{\rm tag}^{\rm X}
-\langle\eta_{\rm tag}^{\rm X} \rangle),\\
\bar{\omega}^{\rm X} (\eta_{\rm tag}^{\rm X} )&=\left(p_{0}^{\rm X} -\frac{\Delta p_{0}^{\rm X}
}{2}\right)+\left(p_{1}^{\rm X} -\frac{\Delta p_{1}^{\rm X} }{2}\right)(\eta_{\rm tag}^{\rm X}
-\langle\eta_{\rm tag}^{\rm X} \rangle),
\end{aligned}
\end{equation}
where $\rm X\in\{OS,SS\}$, $\langle\eta_{\rm tag}^{\rm X} \rangle$ is the mean $\eta_{\rm tag}^{\rm
X}$ of the sample, $p_{0,1}^{\rm X}$ correspond to calibration parameters averaged over \Bs and
\Bsb, and $\Delta p_{0,1}^{\rm X}$ account for \Bs and \Bsb asymmetries in the calibration. Among
other modes, the portability of the SS tagger calibration was checked on $B_s^0\to\phi\phi$ decays~\cite{LHCb-PAPER-2015-056},
which are kinematically similar to the considered signal mode.
The tagging efficiency, $\epsilon_{\rm tag}$, denotes the fraction of candidates with a nonzero
tagging decision. The tagging power of the sample, $\epsilon_{\rm eff}=\epsilon_{\rm
tag}(1-2\langle\omega\rangle)^2$, characterises the tagging performance. Information from the SS
and OS algorithms is combined on a per-event basis (see \eqref{eq:zeta_tagging}) for the
decay-time-dependent amplitude fit discussed in \secref{sec:Model}. The overall effective tagging power is
found to be $(5.15 \pm 0.14)\%$.
The flavour-tagging performance is shown in Table~\ref{tab:flav_tagging}.
When separating the \Bsb and \Bs components at $t=0$, the value of the production asymmetry
$A_p=[\sigma(\Bsb)-\sigma(\Bs)]/[\sigma(\Bsb)+\sigma(\Bs)]$, where $\sigma(\Bs)$\,($\sigma(\Bsb)$) is
the production cross-section for the \Bs(\Bsb) meson, also has to be incorporated in the model.
This asymmetry was measured by LHCb in $pp$ collisions at $\sqrt{s} = 7$ TeV by means of a
decay-time-dependent analysis of $\Bs \to \Dsm \pip$ decays~\cite{LHCb-PAPER-2014-042}. To correct for
the different kinematics of $\Bs \to \Dsm \pip$ and $\BsKpiKpi$ decays, a weighting in bins of \Bs
transverse momentum and pseudorapidity is performed, yielding a value of $A_p=-0.005\pm 0.019$.
No
detection asymmetry need be considered in this analysis since the final state under consideration is charge
symmetric.
\begin{table}
  \centering
  \caption{The flavour-tagging performance of the SS and OS tagging algorithms, as well as the combination of both, for the signal data sample used in the analysis.
  The quoted uncertainty includes both statistical and systematic contributions.}
  \label{tab:flav_tagging}
  \renewcommand{\arraystretch}{1.2}
  \begin{tabular}{  l  c c  }
    Tagging algorithm & $\epsilon_{\rm tag}$ [\%]  & $\epsilon_{\rm eff}$ [\%] \\
    \hline
    SS                & $62.0\pm0.7$  & $1.63\pm0.21$ \\
    OS                & $37.1\pm0.7$  & $3.70\pm0.21$ \\
    Combination       & $75.6\pm0.6$  & $5.15\pm0.14$ \\
  \end{tabular}
\end{table}
\section{Acceptance and resolution effects}
\label{sec:AccepResol}
\begin{figure}
  \includegraphics[width=0.48\textwidth]{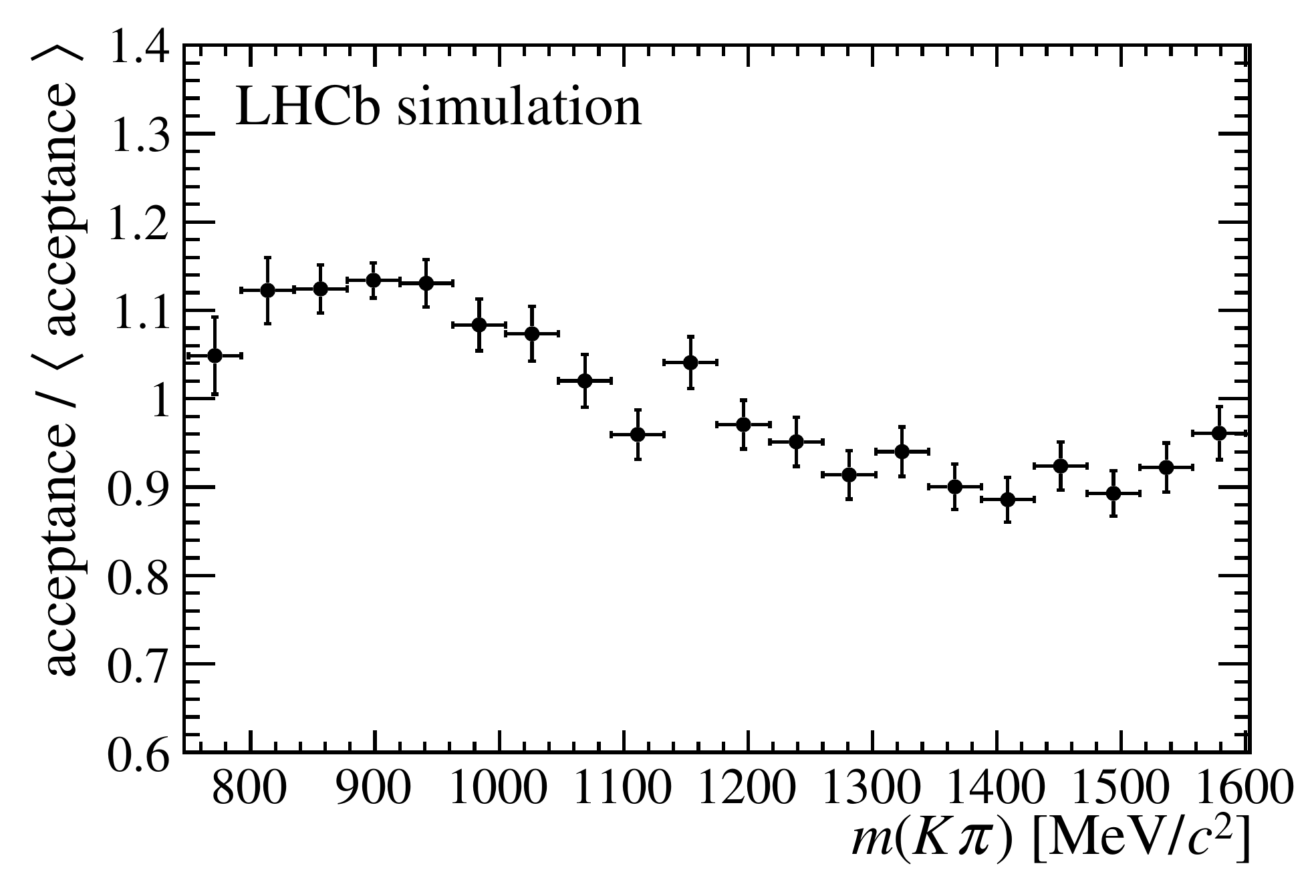}
  \includegraphics[width=0.48\textwidth]{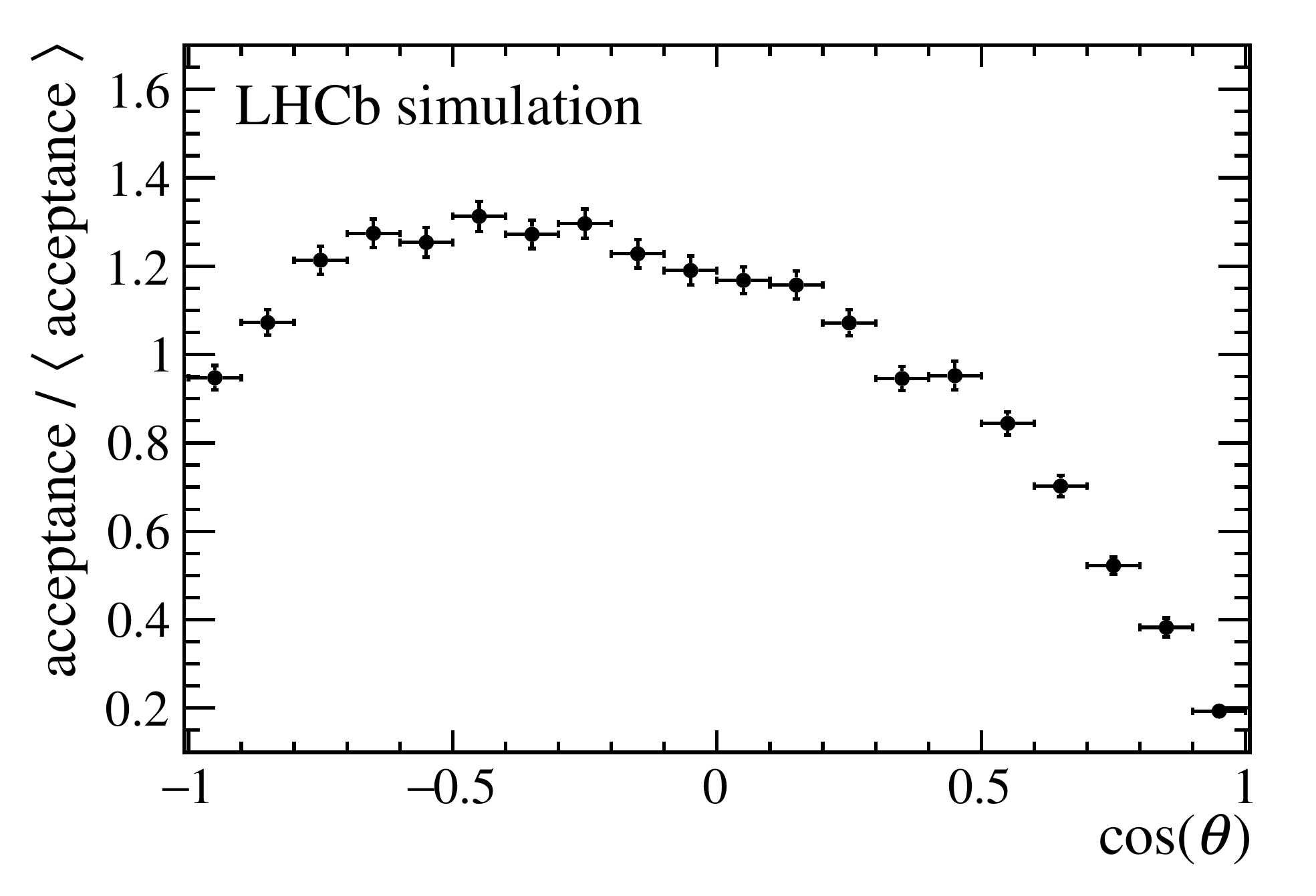} \\
  \includegraphics[width=0.48\textwidth]{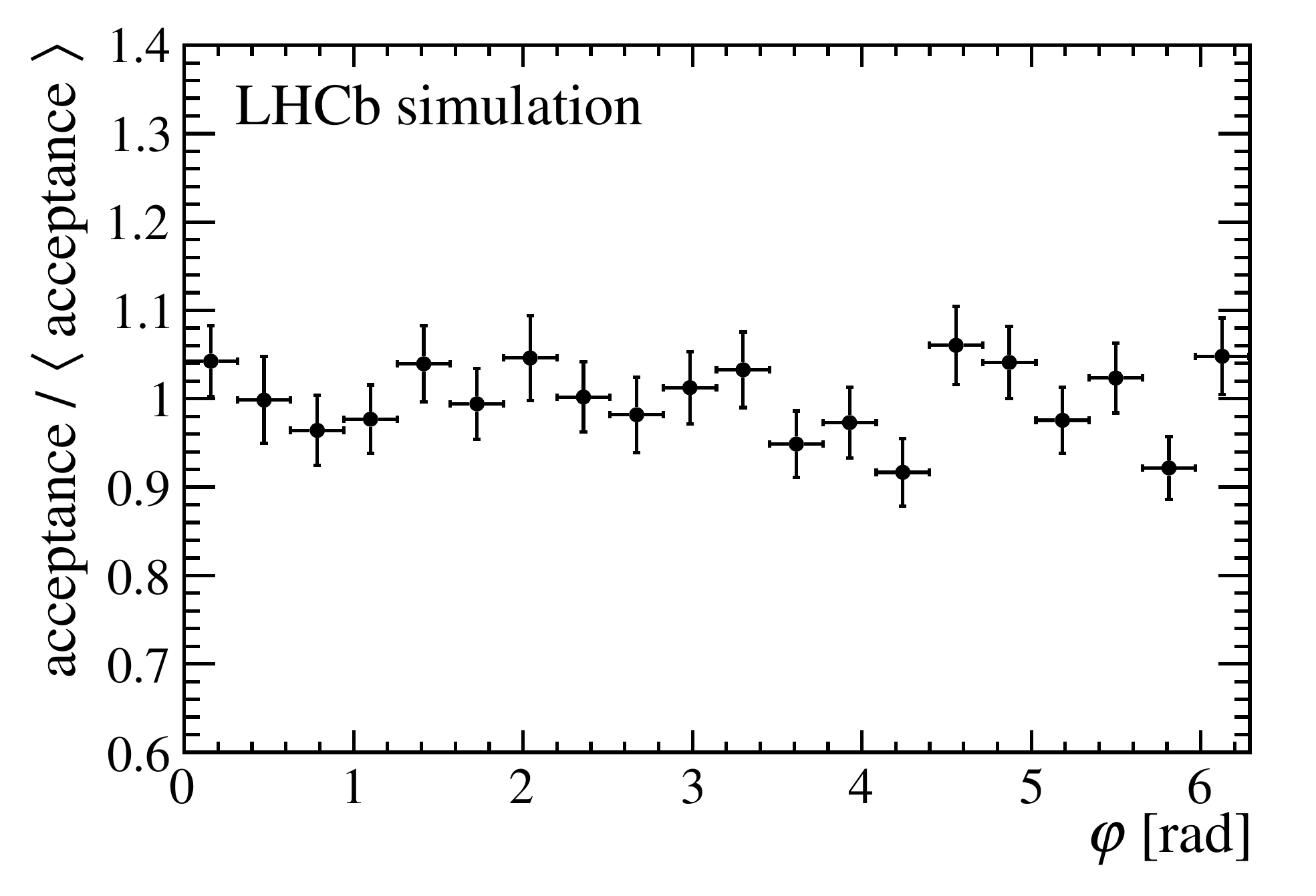}
  \includegraphics[width=0.48\textwidth]{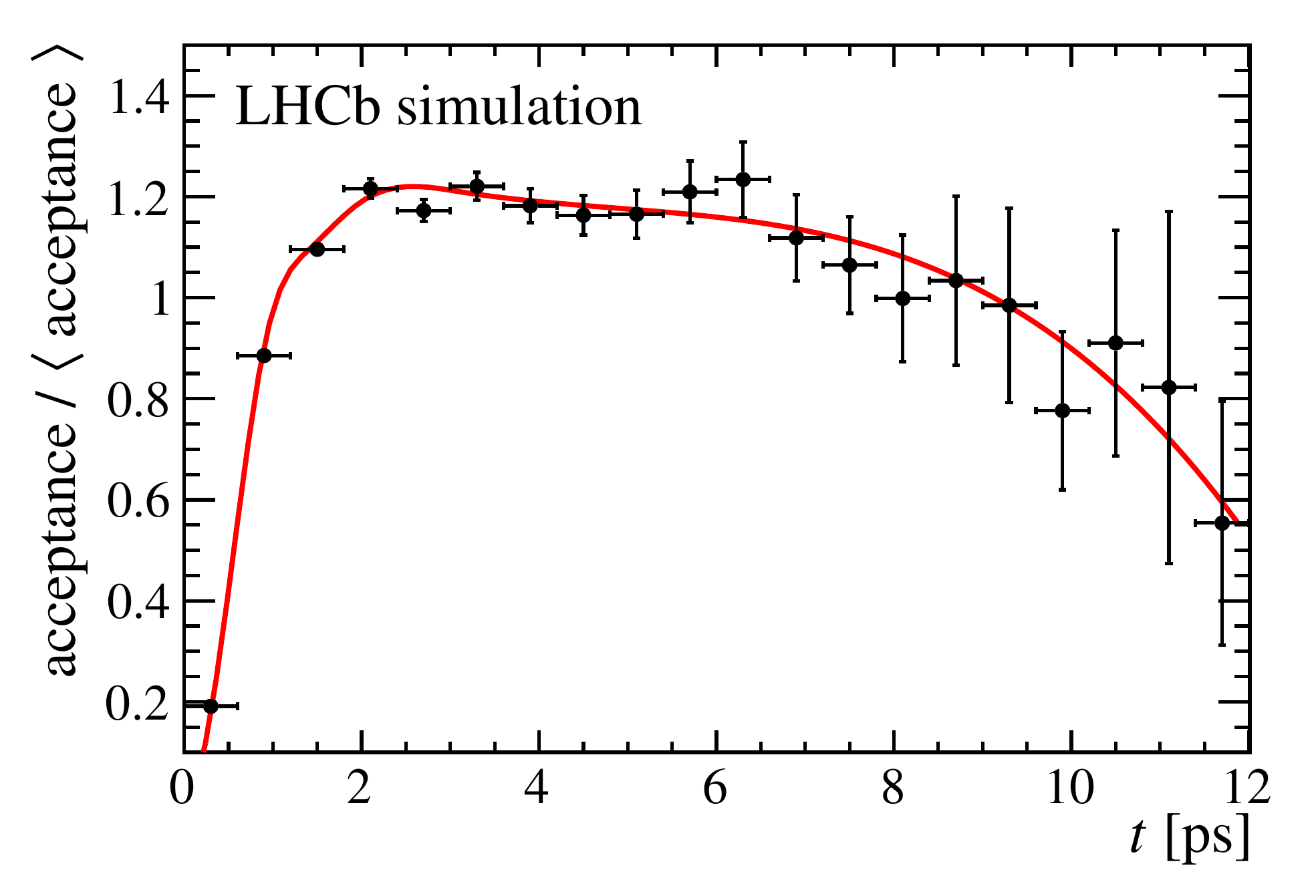} \\
  \caption{Kinematic acceptance and decay-time distributions evaluated with simulated
vector-vector \BsKstKst and pure phase-space \BsKpiKpi candidates scaled by the mean acceptance.
In the bottom right plot the decay-time acceptance obtained from the simulated sample is shown as the black
points and the parametric form of the acceptance obtained with cubic splines is shown as the red
curve. In the other three plots the black points show the acceptance distribution for the masses and
angles. The two $\cos\theta$ variables and the two $m(K\pi)$ masses have been averaged for the
purpose of illustration. In the fit, the kinematic acceptance enters via the normalisation
weights.}
  \label{fig:accep}
\end{figure}

\begin{figure}
  \centering
  \includegraphics[width=0.48\textwidth]{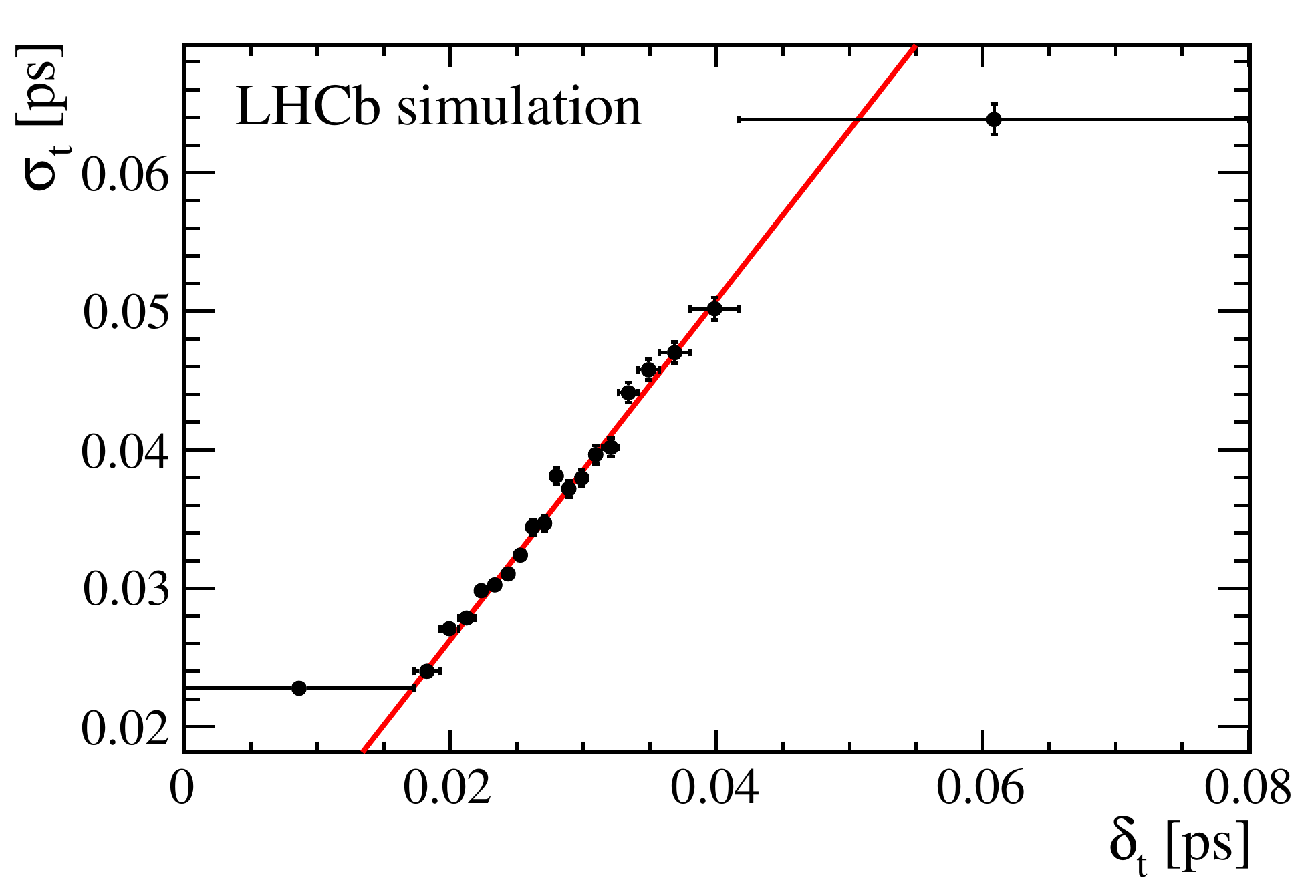}
  \caption{Per-event decay-time resolution, $\sigma_t$, versus the estimated per-event decay-time uncertainty, $\delta_t$,
  obtained from simulated samples containing both vector-vector resonant \BsKstKst and
 phase-space \BsKpiKpiPS events.}
  \label{fig:time_res}
\end{figure}

The LHCb geometrical coverage and selection procedure induce acceptance effects that depend on the
three decay angles, the $K\pi$ two-body invariant masses and the decay time. In addition,
imperfect reconstruction gives rise to resolution effects. Any deviations caused by imperfect
angular and mass resolution are small and are accounted for within the evaluation of systematic
uncertainties (see Sec.~\ref{sec:Systematics}).  However, knowledge of the decay-time resolution is
of key importance in the determination of \phisdd and is consequently included in the
decay-time-dependent fit. In this analysis, both acceptance and resolution effects are studied using
samples of simulated events which have been weighted to match the data distributions in several
important kinematic variables.
In the description of the acceptance, the decay-time-dependent part is factorised with respect to the
part that depends on the kinematic quantities, since they are found to be only $5\%$ correlated.
The acceptance and the decay-time resolutions are determined from simulated events that contain an
appropriate combination of the vector-vector $\BsKstKst$ component with a sample of $\Bs \to
K^+\pi^-K^-\pi^+$ decays generated according to a phase-space distribution. This combination sufficiently populates the phase-space regions to represent the signal decay.
To obtain the acceptance function, the simulated events are weighted by the inverse of the probability density function (PDF)
used for generation (defined in terms of angles, masses and decay time).  The decay-time acceptance is
treated analytically and parameterised using cubic spline functions, following the procedure
outlined in Ref.~\cite{Karbach:2014qba}, with the number of knots chosen to be six. The effect of
this choice is addressed as a systematic uncertainty in Sec.~\ref{sec:Systematics}.  The decay-time
acceptance is shown in Fig.~\ref{fig:accep} (bottom right). The five-dimensional kinematic
acceptance in angles and masses is included by using normalisation weights in the denominator of
the PDF used in the fit to the data, following the procedure
described in Ref.~\cite{CERN-THESIS-2010-124}.  When visualising the fit results (see
Fig.~\ref{fig:fit_result}), the simulated events are weighted using the matrix element of the
amplitude fit model.
For illustrative purposes, some projections of the kinematic acceptance are shown in
Fig.~\ref{fig:accep}.
In order to obtain the best possible sensitivity for the measurement of the \phisdd phase, the time
resolution is evaluated event by event, using the estimated decay-time uncertainty, $\delta_t$,
obtained in the track reconstruction process. This variable is calibrated using the simulation
sample described above to provide the per-event decay-time resolution, $\sigma_t$, using a
linear relationship
\begin{equation}
\label{eq:calsigt}
\sigma_{t}(\delta_t)=p_{0}^{\sigma_t}+p_{1}^{\sigma_t}(\delta_t-\langle\delta_t\rangle),
\end{equation}
where $\langle\delta_t\rangle$ is the mean $\delta_t$ of the sample and $p_{(0,1)}^{\sigma_t}$ are
the calibration parameters. During fitting, $\sigma_t$ is taken to be the width of a Gaussian
resolution function which convolves the decay-time-dependent part of the total amplitude model.
Figure~\ref{fig:time_res} shows the relationship between the estimated decay-time uncertainty,
$\delta_t$, and the calibrated per-event decay-time resolution, $\sigma_t$.
\section{Decay-time-dependent amplitude fit}
\label{sec:Model}
The model used to fit the data is built by taking the squared moduli of the amplitudes
$\braket{f}{\Bs(t)}$ and $\braket{f}{\Bsb(t)}$ introduced in \secref{sec:Phenomenology},
multiplying them by the four-body phase-space factor, incorporating the relevant flavour-tagging
and production-asymmetry parameters, and including the acceptance and resolution factors obtained
in Sec.~\ref{sec:AccepResol}.  The observables $\eta_{\rm tag}^{\rm SS}$, $\eta_{\rm tag}^{\rm OS}$ and
$\delta_t$ (introduced in \secref{sec:Tagging} and \secref{sec:AccepResol}) are treated as
conditional variables.  The effective\footnote{In the PDF used for fitting, the marginal PDFs on
the conditional variables as well as the acceptance function in the numerator are factored out (see
Ref.~\cite{CERN-THESIS-2010-124} for details on the acceptance treatment used in this analysis).}
normalised PDF can be written as
\begin{equation}
\label{eq:fitPDF}
\text{PDF}(t,\Omega)=\frac{\sum_{\alpha=1}^{19}\sum_{\beta\leq \alpha}\Re e[K_{\alpha\beta}(t)
F_{\alpha\beta}(\Omega)]}{\sum_{\alpha'=1}^{19}\sum_{\beta'\leq \alpha'}\Re e[(\int dt'\,K^{\rm untag}_{\alpha'\beta'}(t') \epsilon_{t}(t')) \xi_{\alpha'\beta'}]},
\end{equation}
where the subscript $\alpha$ ($\beta$) represents the state labels
$\{j_1,j_2,h\}$ ($\{j_1',j_2',h'\}$), $K_{\alpha\beta}(t)$ parameterises the decay-time dependence and is defined in
Eq.~(\ref{eq:time_terms}), and $F_{\alpha\beta}(\Omega)$ are terms that parameterise the angular
and mass dependence.  Both the numerator and the denominator of
Eq.~(\ref{eq:fitPDF}) are constructed as a sum over 190 real terms,
which arise when squaring the
amplitudes decomposed in the combination of the nineteen contributing polarisation states.
The decay-time-dependent factors are constructed as
\begin{equation}
\begin{aligned}
K_{\alpha\beta}(t)&=R(t,\delta_t)\otimes \bigg\{e^{-\Gamma_s t} \bigg[\zeta_+ \bigg(a_{\alpha\beta}\cosh\bigg(\frac{1}{2}\Delta\Gamma_s t\bigg)+b_{\alpha\beta}\sinh\bigg(\frac{1}{2}\Delta\Gamma_s t\bigg)\bigg)\\
&+\zeta_- \bigg(c_{\alpha\beta}\cos\left(\Delta m_s t\right)+d_{\alpha\beta}\sin\left(\Delta m_s t\right)\bigg)\bigg]\bigg\},
\label{eq:time_terms}
\end{aligned}
\end{equation}
where $R(t,\delta_t)$ is the decay-time resolution function and the
factors $\zeta_{\pm}$ contain the flavour-tagging and production-asymmetry information. These
factors are
\begin{equation}
\label{eq:zeta_tagging}
  \zeta_{\pm}=\frac{(1+A_p)}{2} P^{\rm OS}(\qtagOS)\, P^{\rm SS}(\qtagSS) \;\pm\;
  \frac{(1-A_p)}{2}\bar{P}^{\rm OS}(\qtagOS)\, \bar{P}^{\rm SS}(\qtagSS),
\end{equation}
where
\begin{equation}
\begin{aligned}
  P^{\rm X}(\qtagT) & = \begin{cases} 1 - \omega^{\rm X}(\eta^{\rm X}) & \text{ for }\qtagT = \phantom{-}1, \\
  1 & \text{ for } \qtagT =\phantom{-}0, \\
  \omega^{\rm X}(\eta^{\rm X}) & \text{ for } \qtagT = -1, \end{cases} \\
  \bar{P}^{\rm X}(\qtagT) & = \begin{cases} \bar{\omega}^{\rm X}(\eta^{\rm X}) & \text{ for }\qtagT = \phantom{-}1, \\
  1 & \text{ for } \qtagT =\phantom{-}0, \\
  1- \bar{\omega}^{\rm X}(\eta^{\rm X}) & \text{ for } \qtagT = -1, \end{cases} \\
\end{aligned}
\end{equation}
with $\rm X\in\{OS,SS\}$.
The complex quantities $a_{\alpha\beta}$, $b_{\alpha\beta}$, $c_{\alpha\beta}$ and $d_{\alpha\beta}$ are defined in terms of the \CP-averaged amplitudes, the \CP-violating parameters and the $\eta_h^{j_1j_2}$ factors, as
\begin{equation}
\begin{aligned}
a_{\alpha\beta} &= \frac{2}{1+|\lambda|^2} \left(A_{\alpha}A_{\beta}^*+\eta_{\alpha} \eta_{\beta} |\lambda|^2A_{\bar{{\alpha}}}A_{\bar{{\beta}}}^*\right),\\
b_{\alpha\beta} &= \frac{-2|\lambda|}{1+|\lambda|^2}\left(\eta_{\beta}
e^{i\phisdd}A_{\alpha}A_{\bar{{\beta}}}^*+\eta_{\alpha} e^{-i\phisdd}A_{\bar{{\alpha}}}A_{\beta}^* \right),\\
c_{\alpha\beta} &=  \frac{2}{1+|\lambda|^2}\left(A_{\alpha}A_{\beta}^*-\eta_{\alpha} \eta_{\beta} |\lambda|^2A_{\bar{{\alpha}}}A_{\bar{{\beta}}}^*\right),\\
d_{\alpha\beta} &= \frac{-2|\lambda|i}{1+|\lambda|^2}\left(\eta_{\beta} e^{i\phisdd}A_{\alpha}A_{\bar{{\beta}}}^*-\eta_{\alpha}
e^{-i\phisdd}A_{\bar{{\alpha}}}A_{\beta}^* \right),
\end{aligned}
\end{equation}
where the bars on the amplitude indices $\alpha$ and $\beta$ denote the \CP transformation of the
considered final state, {\em i.e.}~the change of quantum numbers $j_1j_2\to j_2j_1$. The functions
$K^{\rm untag}_{\alpha\beta}$ are obtained by summing $K_{\alpha\beta}$ over the tagging decisions.
The angular- and mass-dependent terms are constructed as
\begin{equation}
\begin{aligned}
F_{\alpha\beta}(\Omega)&=(2-\delta_{\alpha\beta})\Theta^{j_1 j_2}_{h}(\cos{\theta_1},\cos{\theta_2},\varphi)[\Theta^{j_1' j_2'}_{h'}(\cos{\theta_1},\cos{\theta_2},\varphi)]^*\\
&\times\mathcal{M}_{j_1}(m_1) \mathcal{M}_{j_2}(m_2) \mathcal{M}^*_{j_1'}(m_1) \mathcal{M}^*_{j_2'}(m_2)\\
&\times\mathcal{F}_h^{j_1 j_2}(m_1,m_2) \mathcal{F}_{h'}^{j_1' j_2'}(m_1,m_2) \Phi_4(m_1,m_2),
\label{eq:ang_terms}
\end{aligned}
\end{equation}
where $\delta_{\alpha\beta}$ is the Kronecker delta and the other
terms have been introduced in \secref{sec:Phenomenology}. The decay-time acceptance function,
$\epsilon_t(t)$, and the normalisation weights, $\xi_{\alpha\beta}$, are included in the
denominator of Eq.~(\ref{eq:fitPDF}).  The normalisation weights correspond to angular and
mass integrals that involve the five-dimensional kinematic acceptance, $\epsilon_{\Omega}(\Omega)$,
and are obtained by summing over the events in the simulated sample
\begin{equation}
\xi_{\alpha\beta}\equiv\int{d\Omega\;F_{\alpha\beta}(\Omega)\;\epsilon_{\Omega}(\Omega)}\propto
\sum_{i}^{N_{\rm events}}\frac{F_{\alpha\beta}(\Omega_{i})}{G(\Omega_{i})},
\end{equation}
where $G(\Omega)$ is the model used for generation.
The \CP-conserving amplitudes, $A_h^{j_1,j_2}$, the direct \CP-asymmetry parameter, $|\lambda|$,
and the mixing induced \CP-violating phase, \phisdd, are allowed to vary during the fit. Gaussian
constraints are applied to $\Deltams$, $\Gammas$ and $\DeltaGammas$ from their known
values~\cite{HFLAV16}, and to the flavour-tagging and decay-time resolution calibration parameters,
introduced in \secref{sec:Tagging} and \secref{sec:AccepResol}. The \CP-averaged amplitudes are
characterised in the fit by wave fractions, $f^w$, polarisation fractions, $f^w_h$, and strong
phases, $\delta^w_h$, given by
\begin{equation}
\begin{aligned}
f^w&=\frac{\sum_h |A^w_h|^2}{\sum_{w'}\sum_{h'} |A^{w'}_{h'}|^2},\\
f^w_h&=\frac{|A^w_h|^2}{\sum_{h'} |A^w_{h'}|^2},\\
\delta^w_h&=\arg{(A^w_h)},
\end{aligned}
\end{equation}
with $w$ running over the nine decays under study and $h$ running over the available helicities for
each channel.  With these definitions
it follows that
\begin{equation}
\sum_{w}f^{w}=1,\hspace{1.5cm}\sum_h f_h^{w}=1,\;\forall\,w,
\end{equation}
so not all the fractions are independent of each other, for example $f_{\perp}^{VV}=1-f_{\rm
L}^{VV}-f_{\parallel}^{VV}$. The phase of the longitudinal polarisation amplitude of the
vector-vector component is set to zero to serve as a reference.
\section{Systematic uncertainties}
\label{sec:Systematics}
The decay-time-dependent amplitude model and the fit procedure are cross-checked in several independent
ways: using purely simulated decays, fitting in a narrow window around the dominant
$K^{*0}$ resonance, fitting only in the high-mass region above the $K^{*0}$
resonance, considering higher-spin contributions (whose effect is found to be negligible),
ensuring that there is no bias when repeating the fit procedure on ensembles of pseudo-experiments and
by repeating the fit on subsamples of the data set split by the year of data taking, the magnet polarity
and using a different mass range.
These checks
give compatible results.
Several sources of systematic uncertainty are considered for each of the physical observables
extracted in the decay-time-dependent fit. These are described in this section. A summary of the
systematic uncertainties is given in
Table~\ref{tab:systematics}.
\subsection{Fit to the four-body invariant mass distribution}
The uncertainty on the yield of each of the partially reconstructed components used in the four-body invariant mass fit is
propagated to the decay-time-dependent amplitude fit by recalculating the \sPlot signal weights after
varying each of the yields by one standard deviation. Sources of systematic uncertainty which
arise from mismodelling the shapes of both the background and signal components are calculated by
performing the full fit procedure using alternative parameterisations. The signal is replaced with a
double-sided Crystal Ball function~\cite{Skwarnicki:1986xj} instead of the nominal Ipatia shape
described in Sec.~\ref{sec:Selection} and the combinatorial-background shape is replaced with a
first-order polynomial instead of the nominal exponential function.
\subsection{Weights derived from the \sPlot procedure}
The \sPlot procedure assumes that there is no correlation between the fit variable used to determine the weights,
in this case the four-body invariant mass, $m(\Kp\pim\Km\pip)$, and the projected variables in which the
signal distribution is unfolded, in this case the
three angles and two masses, $\Omega$. This is checked to be valid to a close approximation for signal decays.
In order to assess the impact of any residual correlations in the signal weights, the four-body
mass fit is performed by splitting the data into different bins of $\cos\theta$ for each $(K\pi)$
pair. For each subcategory the four-body fit is repeated and the resulting model is used to
compute a new set of signal weights for the full sample. The largest difference between each
subcategory value and the nominal fit value is taken as the systematic uncertainty.
\subsection{Decay-time-dependent fit procedure}
An ensemble of pseudoexperiments is generated to estimate the bias on the parameters of the
decay-time-dependent fit. For each experiment, a sample with a similar size to the selected signal is
generated using the matrix element of the nominal model (employing the measured amplitudes) and
then refitted to determine the deviation induced in the fit parameters. The systematic uncertainty
is calculated as the mean of the deviation over the ensemble.
\subsection{Decay-time-dependent fit parameterisation}
Several sources of systematic uncertainty originating from the decay-time-dependent fit model have been
studied. These include the parameterisations of the angular momentum centrifugal-barrier factors,
the mean and width of the Breit--Wigner functions and the model for the S-wave propagator. An
alternative model-independent approach is used, as described in \appref{sec:M0}. The systematic uncertainties
are obtained for each of these cases by comparing the fitted parameter values of the alternative model
with the fitted values from the nominal model. Additional contributions from higher mass $(K\pi)$ vector resonances, namely the
$K_1^*(1410)^0$ and the $K_1^*(1680)^0$ states, are also considered. In this case, the size of these
components is first estimated on data through a simplified fit. Afterwards, an ensemble of
pseudoexperiments is generated including these resonances in the model and then refitting with the
nominal PDF. The total systematic uncertainty for the decay-time-dependent fit model is taken as the sum
in quadrature of these alternatives.
\subsection{Acceptance normalisation weights}
The kinematic acceptance weights, explained in Sec.~\ref{sec:Model}, are computed from
simulated samples of limited size, which induces an uncertainty. This systematic uncertainty is calculated using an ensemble of pseudoexperiments in
which the acceptance weights are randomly varied according to their covariance matrix (evaluated on
the simulated sample). The root-mean-square of the distribution of the differences between the
nominal fitted value and the value obtained in each pseudoexperiment is taken as the size of the
systematic uncertainty. This effect is found to be the largest systematic uncertainty impacting the
measurement of the \phisdd phase.
\subsection{Other acceptance and resolution effects}
Various other acceptance and resolution effects for the decay angles, the two $K\pi$ pair masses and
the decay-time are accounted for. Most of these quantities are nominally computed in the
decay-time-dependent fit using simulation samples. Any differences between data and simulation are accounted for by the systematic uncertainties
described in this section. Furthermore, various other effects originating from mismodelling of the decay-time acceptance and
decay-time resolution functions are considered.
Each of these effects are summed in quadrature to provide
the value listed in \tabref{tab:systematics}.
The kinematic and decay-time acceptances, shown in \figref{fig:accep}, are computed from samples of simulated signal events.
Small systematic effects can arise due to differences between the data and the simulated samples. In particular, mismodelling
of the \Bs and the four-track momentum distributions can impact the acceptance in $\cos\theta$. This effect is checked by
producing a data-driven correction for the simulation in several relevant physical quantities.\footnote{The variables used to correct the
distributions of the simulation are the momentum and pseudorapidity of the kaons and pions,
the transverse momentum of the \Bs and the number of tracks in the event.}
This correction is produced using an iterative procedure that removes
any effects arising from differences between the model used in the event generation and the actual decay kinematics of \BsKpiKpi decays.
The systematic uncertainty is computed as the difference in the fit parameters before and after the iterative
correction has been applied.
Systematic effects due to the possible mismodelling of the decay-time-dependent acceptance
are studied by generating ensembles of pseudoexperiments in two different configurations: one in which the decay-time acceptance
spline coefficients are randomised and one in which the configuration of the decay-time acceptance knots is varied.
The nominal decay-time-dependent fit procedure is repeated for each pseudoexperiment and the systematic uncertainty for each of these two effects
is computed as the average deviation of the fit parameters from their generated values over each ensemble.
Sources of systematic uncertainty which affect the decay-time resolution are studied by modifying the
calibration function in \eqref{eq:calsigt} that is used to obtain the per-event decay-time
resolution.  First, the nominal function is substituted by an alternative quadratic form, to asses
the effect of nonlinearity in the calibration. Second, the nominal function is multiplied by a
scale factor that accounts for possible remaining differences between data and simulation. This
scale factor is taken from the analysis of $B_s^0\to J/\psi\phi$ decays performed by LHCb in
Ref.~\cite{LHCb-PAPER-2013-002}. In the both cases, the systematic uncertainties are
obtained by comparing the values resulting from the alternative configurations with the nominal
values.
The effect of the resolution on the masses and angles is studied by generating ensembles of
pseudoexperiments for which the masses and angles are smeared using a multi-dimensional Gaussian
resolution function, obtained from simulation. The systematic uncertainty is computed as the mean deviation
between the fitted and generated values.
\subsection{Production asymmetry}
The uncertainty of the production asymmetry for the \Bs meson is studied by computing the maximum
difference between the nominal conditions and when the production asymmetry is shifted to $\pm
1\sigma$ of its nominal value.
\begin{sidewaystable}
  \begin{center}
    \caption{Summary of the systematic uncertainties on the two \CP parameters, the \CP-averaged fractions and the strong phase differences (in radians) for each of the components listed in Table~\ref{tab:channels}.}
  \label{tab:systematics}
  \end{center}
  \renewcommand{\arraystretch}{1.2}
\resizebox{0.882\textwidth}{!}{
\begin{tabular}{|l|cc|ccccc|cccc|cc|}
\hline
  Parameter & $\phisdd$ [rad] & $|\lambda|$ & $f^{VV}$ & $f_{L}^{VV}$ & $f_{\parallel}^{VV}$ & $\delta_{\parallel}^{VV}$ & $\delta_{\perp}^{VV}$ & $f^{SV}$ & $f^{VS}$ & $\delta^{SV}$ & $\delta^{VS}$ & $f^{SS}$ & $\delta^{SS}$   \\
\hline
Yield and shape of mass model               & $0.012$ & $0.001$ & $0.001$ & $0.004$ & $0.004$ & $0.011$ & $0.020$ & $0.002$ & $0.003$ & $0.023$ & $0.023$ & $0.004$ & $0.012$   \\
Signal weights of mass model                & $0.012$ & $0.007$ & $0.002$ & $0.006$ & $0.005$ & $0.024$ & $0.112$ & $0.004$ & $0.005$ & $0.049$ & $0.022$ & $0.005$ & $0.047$   \\
Decay-time-dependent fit procedure                            & $0.006$ & $0.002$ & $0.001$ & $0.006$ & $0.002$ & $0.007$ & $0.017$ & $0.003$ & $0.002$ & $0.007$ & $0.027$ & $0.001$ & $0.009$   \\
Decay-time-dependent fit parameterisation                      & $0.049$ & $0.013$ & $0.021$ & $0.025$ & $0.026$ & $0.187$ & $0.202$ & $0.042$ & $0.029$ & $0.159$ & $0.234$ & $0.064$ & $0.227$   \\
Acceptance weights (simulated sample size)  & $0.106$ & $0.078$ & $0.004$ & $0.031$ & $0.029$ & $0.236$ & $0.564$ & $0.037$ & $0.039$ & $0.250$ & $0.290$ & $0.015$ & $0.256$   \\
Other acceptance and resolution effects     & $0.063$ & $0.008$ & $0.005$ & $0.018$ & $0.005$ & $0.136$ & $0.149$ & $0.006$ & $0.004$ & $0.167$ & $0.124$ & $0.017$ & $0.194$   \\
Production asymmetry                        & $0.002$ & $0.002$ & $0.000$ & $0.000$ & $0.000$ & $0.001$ & $0.017$ & $0.002$ & $0.002$ & $0.002$ & $0.008$ & $0.000$ & $0.002$   \\
\hline
Total                                       & $0.141$ & $0.089$ & $0.024$ & $0.046$ & $0.042$ & $0.333$ & $0.641$ & $0.071$ & $0.065$ & $0.346$ & $0.405$ & $0.069$ & $0.399$   \\
\hline
\end{tabular}}
\resizebox{\textwidth}{!}{
\begin{tabular}{|l|cccc|cccccccccccc|}
\hline
  Parameter                               & $f^{ST}$ & $f^{TS}$ & $\delta^{ST}$ & $\delta^{TS}$ & $f^{VT}$ & $f_{L}^{VT}$ & $f_{\parallel}^{VT}$ & $f^{TV}$ & $f_{L}^{TV}$ & $f_{\parallel}^{TV}$ & $\delta_{0}^{VT}$ & $\delta_{\parallel}^{VT}$ & $\delta_{\perp}^{VT}$ & $\delta_{0}^{TV}$ & $\delta_{\parallel}^{TV}$ & $\delta_{\perp}^{TV}$    \\
\hline
Yield and shape of mass model               & $0.002$ & $0.004$ & $0.111$ & $0.023 $ & $0.001$ & $0.003$ & $0.001$ & $0.001$ & $0.043$ & $0.025$ & $0.023$ & $0.055$ & $0.110$ & $0.053$ & $0.018$ & $0.065$  \\
Signal weights of mass model                & $0.004$ & $0.006$ & $0.151$ & $0.105 $ & $0.002$ & $0.003$ & $0.001$ & $0.001$ & $0.043$ & $0.029$ & $0.025$ & $0.131$ & $0.126$ & $0.080$ & $0.073$ & $0.150$  \\
Decay-time-dependent fit procedure                            & $0.001$ & $0.002$ & $0.248$ & $0.017 $ & $0.002$ & $0.004$ & $0.002$ & $0.002$ & $0.008$ & $0.005$ & $0.012$ & $0.069$ & $0.025$ & $0.062$ & $0.017$ & $0.030$  \\
Decay-time-dependent fit parameterisation                      & $0.006$ & $0.017$ & $0.736$ & $0.247 $ & $0.011$ & $0.053$ & $0.019$ & $0.008$ & $0.080$ & $0.048$ & $0.286$ & $0.308$ & $0.260$ & $0.260$ & $0.228$ & $0.405$  \\
Acceptance weights (simulated sample size)  & $0.014$ & $0.015$ & $1.463$ & $0.719 $ & $0.026$ & $0.145$ & $0.054$ & $0.027$ & $0.199$ & $0.102$ & $1.117$ & $1.080$ & $0.888$ & $0.712$ & $0.417$ & $0.947$  \\
Other acceptance and resolution effects     & $0.002$ & $0.003$ & $0.184$ & $0.226 $ & $0.015$ & $0.024$ & $0.004$ & $0.005$ & $0.045$ & $0.017$ & $0.163$ & $0.168$ & $0.191$ & $0.229$ & $0.246$ & $0.171$  \\
Production asymmetry                        & $0.001$ & $0.001$ & $0.037$ & $0.026 $ & $0.001$ & $0.003$ & $0.001$ & $0.002$ & $0.012$ & $0.006$ & $0.015$ & $0.030$ & $0.018$ & $0.003$ & $0.007$ & $0.041$  \\
\hline
Total                                       & $0.031$ & $0.033$ & $1.688$ & $0.817 $ & $0.049$ & $0.165$ & $0.063$ & $0.048$ & $0.252$ & $0.143$ & $1.171$ & $1.159$ & $0.970$ & $0.802$ & $0.546$ & $1.076$  \\
\hline
\end{tabular}
  }
\resizebox{0.731\textwidth}{!}{
\begin{tabular}{|l|cccccccccc|}
\hline
  Parameter                                & $f^{TT}$ & $f_{L}^{TT}$ & $f_{\parallel_1}^{TT}$ & $f_{\perp_1}^{TT}$ & $f_{\parallel_2}^{TT}$ & $\delta_{0}^{TT}$ & $\delta_{\parallel_1}^{TT}$ & $\delta_{\perp_1}^{TT}$ & $\delta_{\parallel_2}^{TT}$ & $\delta_{\perp_2}^{TT}$   \\
\hline
Yield and shape of mass model              & $0.000$ & $0.045$ & $0.019$ & $0.037$ & $0.002$ & $0.038$ & $0.027$ & $0.009$ & $0.079$ & $0.114$   \\
Signal weights of mass model               & $0.000$ & $0.066$ & $0.025$ & $0.024$ & $0.002$ & $0.147$ & $0.046$ & $0.112$ & $0.123$ & $0.215$   \\
Decay-time-dependent fit procedure                           & $0.001$ & $0.022$ & $0.022$ & $0.014$ & $0.004$ & $0.127$ & $0.036$ & $0.068$ & $0.058$ & $0.040$   \\
Decay-time-dependent fit parameterisation                     & $0.005$ & $0.051$ & $0.071$ & $0.113$ & $0.038$ & $1.213$ & $0.199$ & $0.685$ & $0.820$ & $0.476$   \\
Acceptance weights (simulated sample size) & $0.003$ & $0.135$ & $0.110$ & $0.127$ & $0.077$ & $1.328$ & $0.454$ & $1.348$ & $1.443$ & $1.161$   \\
Other acceptance and resolution effects    & $0.002$ & $0.031$ & $0.028$ & $0.056$ & $0.024$ & $0.226$ & $0.275$ & $0.156$ & $0.343$ & $0.301$   \\
Production asymmetry                       & $0.000$ & $0.002$ & $0.001$ & $0.008$ & $0.003$ & $0.005$ & $0.002$ & $0.062$ & $0.015$ & $0.043$   \\
\hline
Total                                      & $0.007$ & $0.176$ & $0.142$ & $0.205$ & $0.107$ & $1.825$ & $0.573$ & $1.546$ & $1.706$ & $1.330$   \\
\hline
\end{tabular}
  }
\end{sidewaystable}
\section{Fit results}
\label{sec:Results}
\begin{figure}
\centering
  \includegraphics[width=0.48\textwidth]{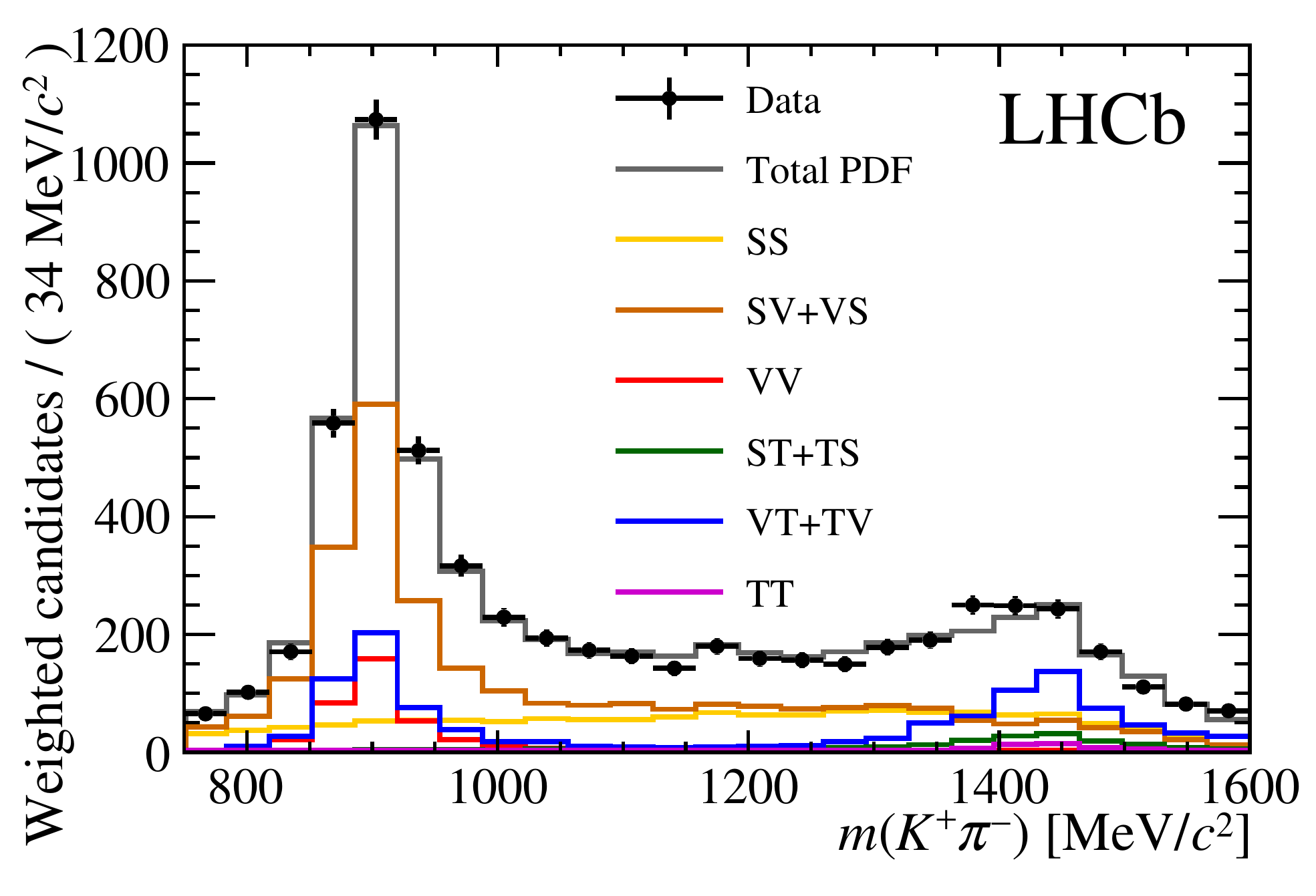}
  \includegraphics[width=0.48\textwidth]{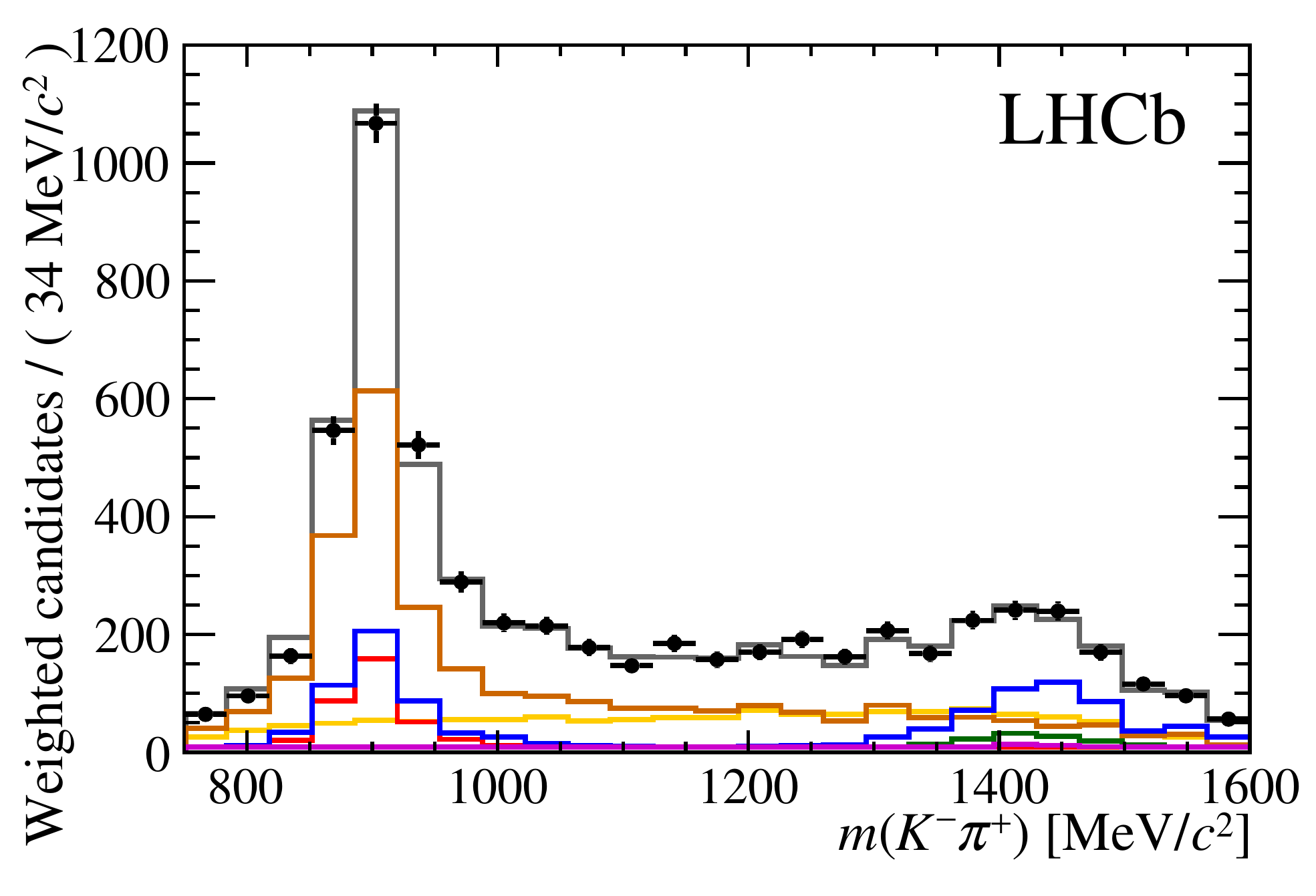} \\
  \includegraphics[width=0.48\textwidth]{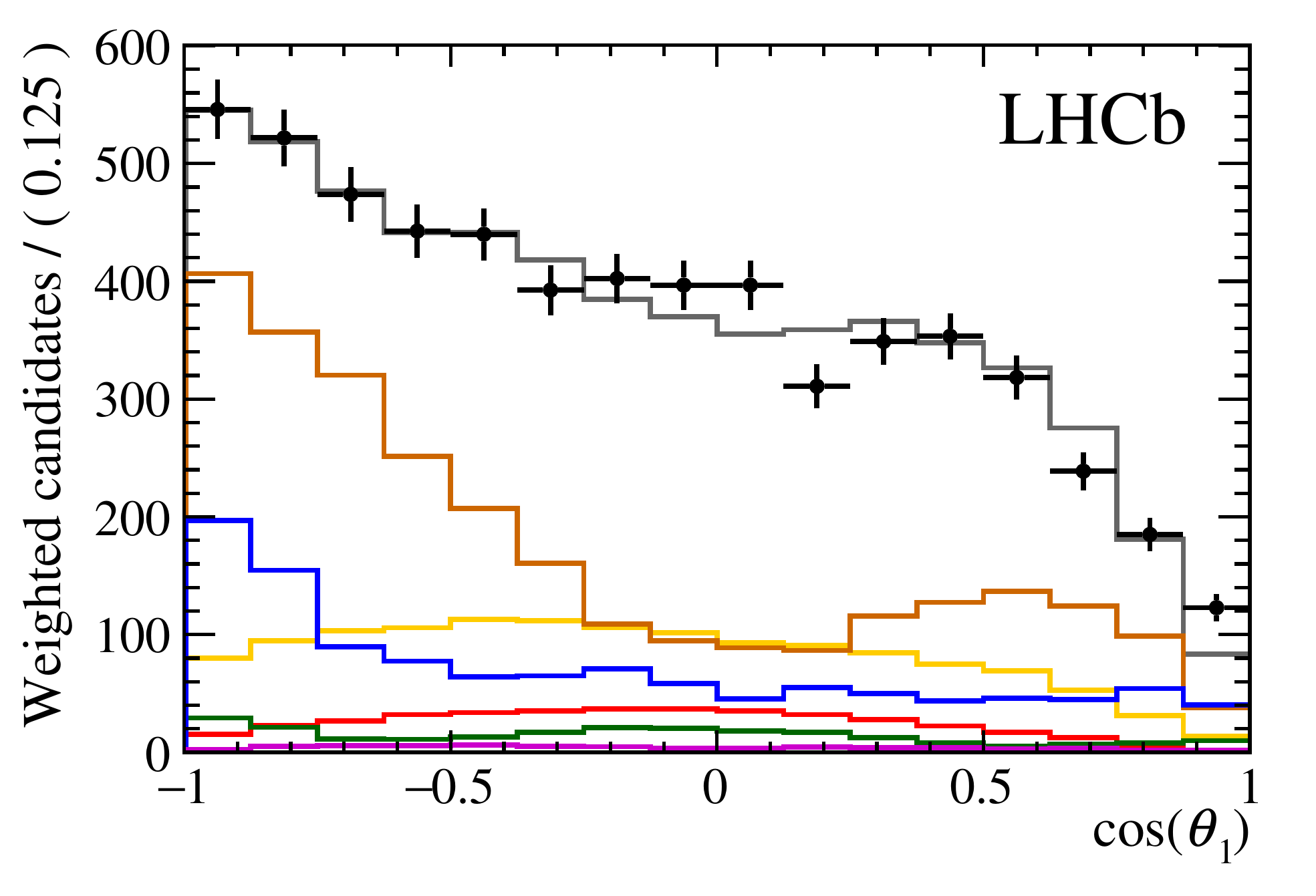}
  \includegraphics[width=0.48\textwidth]{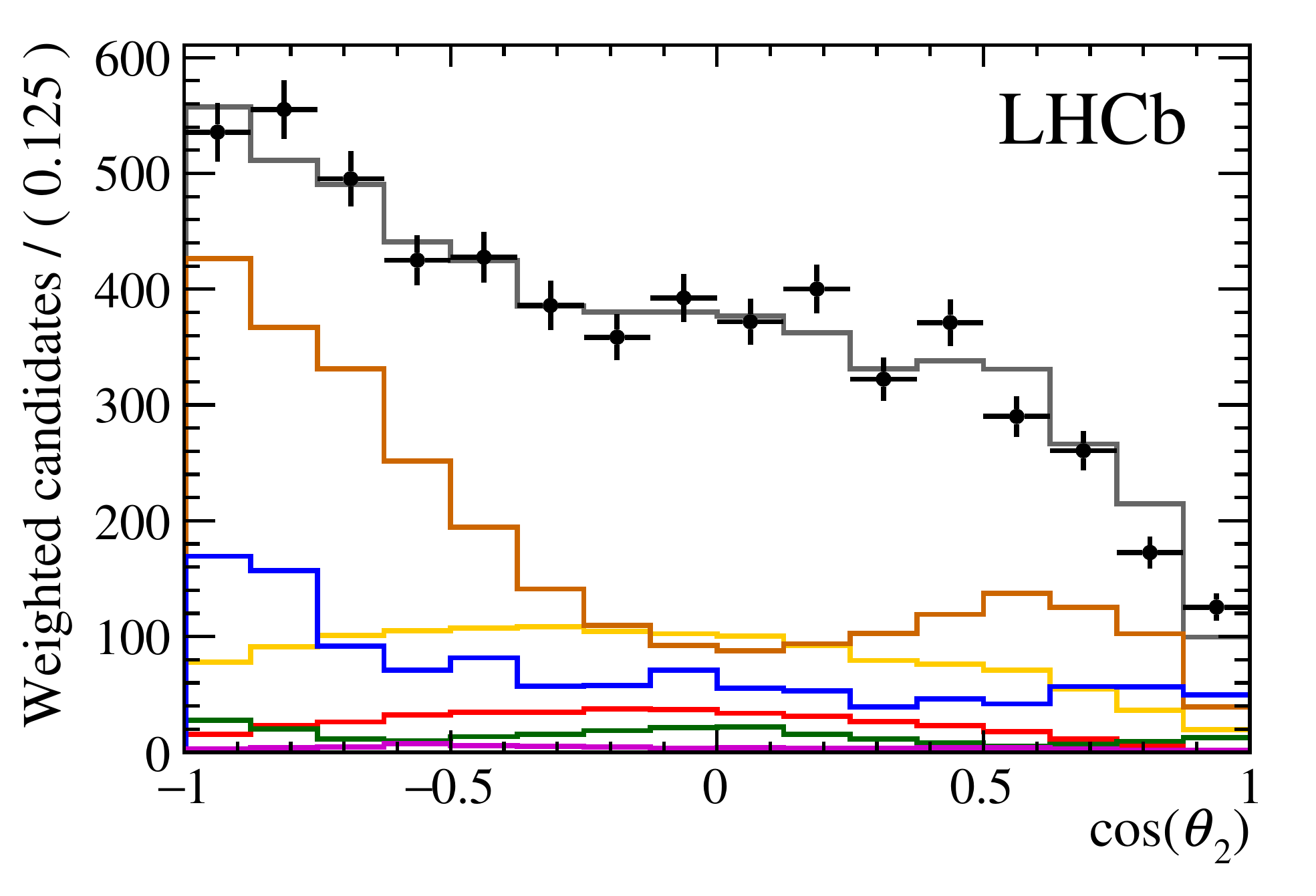} \\
  \includegraphics[width=0.48\textwidth]{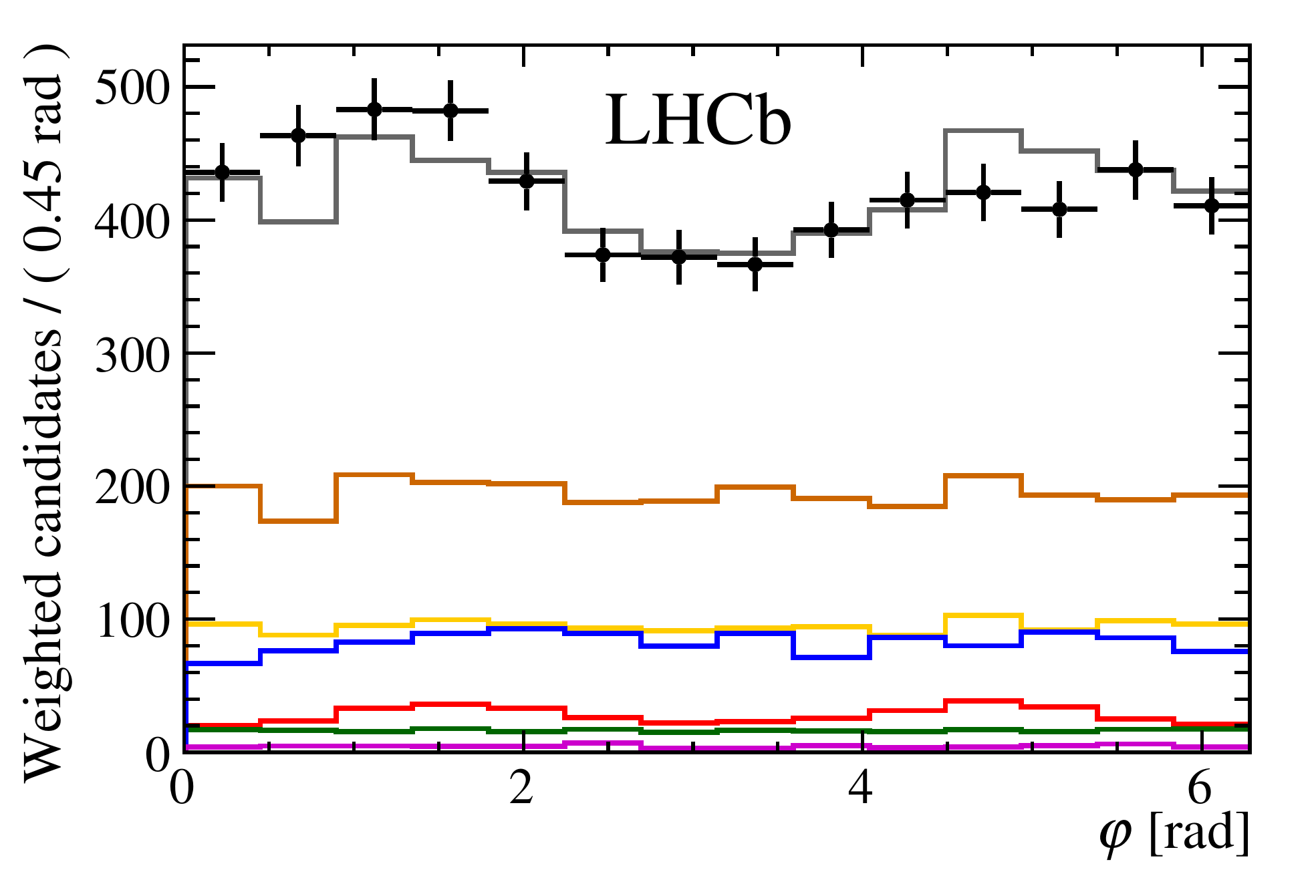}
  \includegraphics[width=0.48\textwidth]{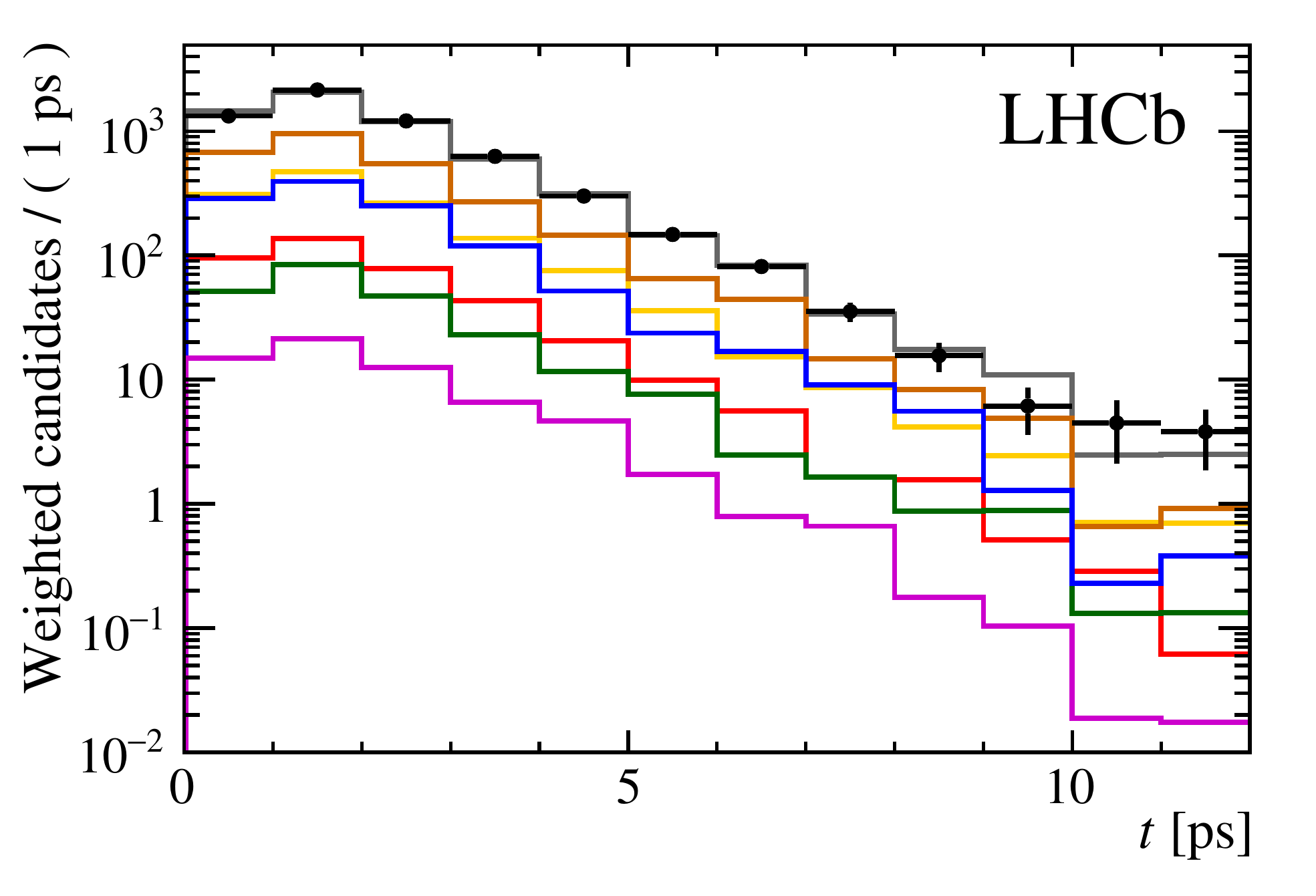}
  \caption{One-dimensional projections of the decay-time-dependent, flavour-tagged fit to (black points) the \sPlot weighted data for (top row) the two $(K\pi)$ invariant masses, (middle row) the two $(K\pi)$ decay plane angles, (bottom left) the angle between the two $(K,\pi)$ decay planes and (bottom right) the decay-time. The solid gray line represents the total fit model along with the \CP-averaged components for each
  contributing decay.}
  \label{fig:fit_result}
\end{figure}

An unbinned maximum likelihood fit is applied to the background-subtracted data using the PDF defined in Eq.~(\ref{eq:fitPDF}). The large computational load due to the complexity of the fit
motivates the parallelisation of the process on a Graphics Processing Unit (GPU), for which the
\textsc{Ipanema} software package~\cite{Santos:2017dqe,kloeckner_pycuda_2012} is used.
The one-dimensional projections of the results in the six analysis variables are shown in
Fig.~\ref{fig:fit_result} along with the separate components from the contributing decay modes
listed in Table~\ref{tab:channels}. The resulting fit values for the common \CP observables,
\phisdd and $|\lambda|$, as well as the \CP-averaged fractions, $f_i$, and polarisation strong-phase
differences, $\delta_i$, for each component are given in Table~\ref{tab:fit_params}. The central
values are given along with the statistical uncertainties obtained from the fit and the systematic
uncertainties, which are discussed in Sec.~\ref{sec:Systematics}.  These are the first measurements in a \btoddbs transition
of the \CP-violation parameter $|\lambda| = 1.035 \pm 0.034 \pm 0.089$ and the \CP-violating weak
phase $\phisdd = -0.10 \pm 0.13 \pm 0.14\rad$. Both are consistent with
no \CP violation and with the SM predictions.
In the region of phase space considered, the \BsKstKst vector-vector component has a relatively
small fraction, of $f^{VV}=0.067 \pm 0.004 \pm 0.024$, mainly due to the large scalar $K\pi$
contributions. Indeed, a relatively large contribution from the scalar-scalar double \swave
fraction is determined to be $f^{SS}=0.225 \pm 0.010 \pm 0.069$. The tensor-tensor double \dwave
fraction is measured to be $f^{TT}=0.011 \pm 0.003 \pm 0.007$. The cross-term contributions from
the scalar with the vector combination (single \swave) and the vector with the tensor combination (single \dwave) are
also found to be large, $f^{SV}=0.329 \pm 0.015 \pm 0.071$, $f^{VS}=0.133 \pm 0.013 \pm 0.065$,
$f^{VT}=0.160 \pm 0.016 \pm 0.049$ and $f^{TV}=0.036 \pm 0.014 \pm 0.048$, while a small
contribution from the scalar with the tensor combination is found, $f^{TS}=0.025 \pm 0.007 \pm 0.033$ and
$f^{ST}=0.014 \pm 0.006 \pm 0.031$. The values of the longitudinal polarisation fractions of
the vector-vector and tensor-tensor components are found to be small, $f_{\rm L}^{TT}=0.25 \pm 0.14
\pm 0.18$ and $f_{\rm L}^{VV}=0.208\pm0.032\pm0.046$, while the longitudinal polarisation fractions
of the vector with the tensor components are measured to be large, $f_{\rm L}^{VT}=0.911 \pm 0.020 \pm
0.165$ and $f_{\rm L}^{TV}=0.62 \pm 0.16 \pm 0.25$.
\begin{table}
  \begin{center}
    \caption{Results of the decay-time-dependent amplitude fit to data. The first uncertainty is statistical and the second uncertainty is systematic.}
  \renewcommand{\arraystretch}{1.2}
  \begin{tabular}{|c|c|}
  \hline
  Parameter & Value \\
  \hline
  \hline
  \multicolumn{2}{|c|}{Common parameters}\\
  \hline
  $\phisdd$ [rad]             & $-0.10\phantom{0} \pm 0.13\phantom{0} \pm 0.14\phantom{0}$ \\
  $|\lambda|$                 & $\phantom{-}1.035  \pm 0.034 \pm 0.089$ \\
  \hline
  \hline
  \multicolumn{2}{|c|}{Vector/Vector (VV)}\\
  \hline
  $f^{VV}$                        & $\phantom{-}0.067 \pm 0.004 \pm 0.024$ \\
  $f_{\rm L}^{VV}$                & $\phantom{-}0.208 \pm 0.032 \pm 0.046$ \\
  $f_{\parallel}^{VV}$            & $\phantom{-}0.297 \pm 0.029 \pm 0.042$ \\
  $\delta^{VV}_{\parallel}$ [rad] & $\phantom{-}2.40\phantom{0} \pm 0.11\phantom{0} \pm 0.33\phantom{0}$ \\
  $\delta^{VV}_{\perp}$ [rad]     & $\phantom{-}2.62\phantom{0} \pm 0.26\phantom{0} \pm 0.64\phantom{0}$ \\
  \hline
  \hline
  \multicolumn{2}{|c|}{Scalar/Vector (SV and VS)}\\
  \hline
  $f^{SV}$      & $\phantom{-}0.329 \pm 0.015 \pm 0.071$ \\
  $f^{VS}$      & $\phantom{-}0.133 \pm 0.013 \pm 0.065$ \\
  $\delta^{SV}$ [rad] & $-1.31\phantom{0} \pm 0.10\phantom{0} \pm 0.35\phantom{0}$ \\
  $\delta^{VS}$ [rad] & $\phantom{-}1.86\phantom{0} \pm 0.11\phantom{0} \pm 0.41\phantom{0}$ \\
  \hline
  \hline
  \multicolumn{2}{|c|}{Scalar/Scalar (SS)}\\
  \hline
  $f^{SS}$      & $\phantom{-}0.225 \pm 0.010 \pm 0.069$ \\
  $\delta^{SS}$ [rad] & $\phantom{-}1.07\phantom{0} \pm 0.10\phantom{0} \pm 0.40\phantom{0}$ \\
  \hline
  \hline
    \multicolumn{2}{|c|}{Scalar/Tensor (ST and TS)}\\
  \hline
  $f^{ST}$      & $\phantom{-}0.014 \pm 0.006 \pm 0.031$ \\
  $f^{TS}$      & $\phantom{-}0.025 \pm 0.007 \pm 0.033$ \\
  $\delta^{ST}$ [rad] & $-2.3\phantom{00} \pm 0.4\phantom{00} \pm 1.7\phantom{00}$ \\
  $\delta^{TS}$ [rad] & $-0.10\phantom{0} \pm 0.26\phantom{0} \pm 0.82\phantom{0}$ \\
  \hline
  \multicolumn{2}{c}{\vspace{0.32cm}} \\
  \end{tabular}
  \begin{tabular}{|c|c|}
  \hline
  Parameter & Value \\
  \hline
  \hline
  \multicolumn{2}{|c|}{Vector/Tensor (VT and TV)}\\
  \hline
  $f^{VT}$                        & $\phantom{-}0.160 \pm 0.016 \pm 0.049$ \\
  $f_{\rm L}^{VT}$                & $\phantom{-}0.911 \pm 0.020 \pm 0.165$ \\
  $f_{\parallel}^{VT}$            & $\phantom{-}0.012 \pm 0.008 \pm 0.053$ \\
  $f^{TV}$                        & $\phantom{-}0.036 \pm 0.014 \pm 0.048$ \\
  $f_{\rm L}^{TV}$                & $\phantom{-}0.62\phantom{0} \pm 0.16\phantom{0} \pm 0.25\phantom{0}$ \\
  $f_{\parallel}^{TV}$            & $\phantom{-}0.24\phantom{0} \pm 0.10\phantom{0} \pm 0.14\phantom{0}$ \\
  $\delta^{VT}_{0}$ [rad]         & $-2.06\phantom{0} \pm 0.19\phantom{0} \pm 1.17\phantom{0}$ \\
  $\delta^{VT}_{\parallel}$ [rad] & $-1.8\phantom{00} \pm 0.4\phantom{00} \pm 1.0\phantom{00}$ \\
  $\delta^{VT}_{\perp}$ [rad]     & $-3.2\phantom{00} \pm 0.3\phantom{00} \pm 1.2\phantom{00}$ \\
  $\delta^{TV}_{0}$ [rad]         & $\phantom{-}1.91\phantom{0} \pm 0.30\phantom{0} \pm 0.80\phantom{0}$ \\
  $\delta^{TV}_{\parallel}$ [rad] & $\phantom{-}1.09\phantom{0} \pm 0.19\phantom{0} \pm 0.55\phantom{0}$ \\
  $\delta^{TV}_{\perp}$ [rad]     & $\phantom{-}0.2\phantom{00} \pm 0.4\phantom{00} \pm 1.1\phantom{00}$ \\
  \hline
  \hline
  \multicolumn{2}{|c|}{Tensor/Tensor (TT)}\\
  \hline
  $f^{TT}$                   & $\phantom{-}0.011 \pm 0.003 \pm 0.007$ \\
  $f_{\rm L}^{TT}$                 & $\phantom{-}0.25\phantom{0} \pm 0.14\phantom{0} \pm 0.18\phantom{0}$ \\
  $f_{\parallel_1}^{TT}$      & $\phantom{-}0.17\phantom{0} \pm 0.11\phantom{0} \pm 0.14\phantom{0}$ \\
  $f_{\perp_1}^{TT}$          & $\phantom{-}0.30\phantom{0} \pm 0.18\phantom{0} \pm 0.21\phantom{0}$ \\
  $f_{\parallel_2}^{TT}$      & $\phantom{-}0.015 \pm 0.033 \pm 0.107$ \\
  $\delta^{TT}_{0}$ [rad]          & $\phantom{-}1.3\phantom{00} \pm 0.5\phantom{00} \pm 1.8\phantom{00}$ \\
  $\delta^{TT}_{\parallel_1}$ [rad] & $\phantom{-}3.00\phantom{0} \pm 0.29\phantom{0} \pm 0.57\phantom{0}$ \\
  $\delta^{TT}_{\perp_1}$ [rad]     & $\phantom{-}2.6\phantom{00} \pm 0.4\phantom{00} \pm 1.5\phantom{00}$ \\
  $\delta^{TT}_{\parallel_2}$ [rad] & $\phantom{-}2.3\phantom{00} \pm 0.8\phantom{00} \pm 1.7\phantom{00}$ \\
  $\delta^{TT}_{\perp_2}$ [rad]     & $\phantom{-}0.7\phantom{00} \pm 0.6\phantom{00} \pm 1.3\phantom{00}$ \\
  \hline
  \end{tabular}
  \label{tab:fit_params}
  \end{center}
\end{table}
\section{Summary}
\label{sec:Summary}
A flavour-tagged decay-time-dependent amplitude analysis of the $\Bs\to(\Kp\pim)(\Km\pip)$ decay, for
$(\Kpm\pimp)$ invariant masses in the range from 750 to 1600\mevcc, is performed on a data set
corresponding to an integrated luminosity of $3.0\invfb$ obtained by the LHCb experiment with
$pp$ collisions at $\sqrt{s}=7\tev$ and $\sqrt{s}=8\tev$. Several quasi-two-body decay components
are considered, corresponding to $(\Kpm\pimp)$ combinations with spins of 0, 1 and 2. The
longitudinal polarisation fraction for the \BsKstKst vector-vector decay is determined to be
$f_{\rm L}^{VV}=0.208\pm0.032\pm0.046$, where the first uncertainty is statistical and the second
one systematic. This confirms, with improved precision, the relatively low value reported
previously by LHCb~\cite{LHCb-PAPER-2014-068}. The first determination of the \CP asymmetry of the
$(K^+\pi^-)(K^-\pi^+)$ final state and the best, sometimes the first, measurements of 19
\CP-averaged amplitude parameters corresponding to scalar, vector and tensor final states, are also
reported. This analysis determines for the first time the mixing-induced \CP-violating phase
$\phi_s$ using a \btoddbs transition. The value of this phase is measured to be $\phisdd = -0.10
\pm 0.13 \pm 0.14\rad$, which is consistent with both the SM expectation~\cite{Bhattacharya:2013sga}
and the corresponding LHCb result of $\phisss= -0.17 \pm 0.15 \pm 0.03\rad$ measured using $\Bs\to\phi\phi$ decays~\cite{LHCb-PAPER-2014-026}.
The statistical uncertainty of the two measurements is at a similar level although the systematic uncertainty of this
measurement is larger, which is mainly due to the
treatment of the multi-dimensional acceptance.
It is expected that this can be reduced by increasing the size of the simulation sample used to determine the acceptance effects.
Most other sources of systematic uncertainty are expected to scale with larger data samples.

\section*{Acknowledgements}
\noindent We express our gratitude to our colleagues in the CERN
accelerator departments for the excellent performance of the LHC. We
thank the technical and administrative staff at the LHCb
institutes. We acknowledge support from CERN and from the national
agencies: CAPES, CNPq, FAPERJ and FINEP (Brazil); MOST and NSFC
(China); CNRS/IN2P3 (France); BMBF, DFG and MPG (Germany); INFN
(Italy); NWO (The Netherlands); MNiSW and NCN (Poland); MEN/IFA
(Romania); MinES and FASO (Russia); MinECo (Spain); SNSF and SER
(Switzerland); NASU (Ukraine); STFC (United Kingdom); NSF (USA).  We
acknowledge the computing resources that are provided by CERN, IN2P3
(France), KIT and DESY (Germany), INFN (Italy), SURF (The
Netherlands), PIC (Spain), GridPP (United Kingdom), RRCKI and Yandex
LLC (Russia), CSCS (Switzerland), IFIN-HH (Romania), CBPF (Brazil),
PL-GRID (Poland) and OSC (USA). We are indebted to the communities
behind the multiple open-source software packages on which we depend.
Individual groups or members have received support from AvH Foundation
(Germany), EPLANET, Marie Sk\l{}odowska-Curie Actions and ERC
(European Union), ANR, Labex P2IO and OCEVU, and R\'{e}gion
Auvergne-Rh\^{o}ne-Alpes (France), RFBR, RSF and Yandex LLC (Russia),
GVA, XuntaGal and GENCAT (Spain), Herchel Smith Fund, the Royal
Society, the English-Speaking Union and the Leverhulme Trust (United
Kingdom).

\clearpage
{\noindent\normalfont\bfseries\Large Appendices}
\appendix
\section{Angular distributions}
\label{sec:angular}
The angular dependence of the decay amplitudes introduced in \eqref{eq:angmassamplitudes} is shown in \tabref{tab:ang_amplitudes}.
\begin{table}[h]
\begin{center}
\caption{Functions containing the angular dependence of the amplitudes, as introduced in \eqref{eq:angmassamplitudes}. For a discussion on some of the angular terms see Ref.~\cite{Bhattacharya:2013sga}.}
\begin{tabular}{|l|l|r|l|l|}\hline
 $j_1$ & $j_2$ & $h$ & $Y_{\ell_1}^{m_1}(\theta_1,-\varphi)Y_{\ell_2}^{m_2}(\pi-\theta_2,0)$
  & $\Theta^{j_1j_2}_{h}(\cos\theta_1,\cos\theta_2,\varphi)$         \\ [1mm] \hline\hline
 0 & 0 & ${0          }$ & $ \sqrt{\pi} Y_0^0 Y_0^0  $                  & $\frac{1}{2\sqrt{2\pi}}$ \\ [1mm] \hline
 0 & 1 & ${0          }$ & $ \sqrt{\pi} Y_0^0 Y_1^0  $                  & $-\frac{\sqrt{3}}{2\sqrt{2\pi}}\cos\theta_2$ \\ [1mm] \hline
 1 & 0 & ${0          }$ & $ \sqrt{\pi} Y_1^0 Y_0^0  $                  & $\frac{\sqrt{3}}{2\sqrt{2\pi}}\cos\theta_1$ \\ [1mm] \hline
 0 & 2 & ${0          }$ & $ \sqrt{\pi} Y_0^0 Y_2^0  $                  & $\frac{\sqrt{5}}{4\sqrt{2\pi}}(3\cos^2\theta_2-1)$ \\ [1mm] \hline
 2 & 0 & ${0          }$ & $ \sqrt{\pi} Y_2^0 Y_0^0  $                  & $\frac{\sqrt{5}}{4\sqrt{2\pi}}(3\cos^2\theta_1-1)$ \\ [1mm] \hline
 1 & 1 & ${0          }$ & $ \sqrt{\pi} Y_1^0 Y_1^0  $                  & $-\frac{3}{2\sqrt{2\pi}} \cos\theta_1\cos\theta_2$ \\ [1mm]
 1 & 1 & ${\parallel  }$ & $ \frac{\sqrt{\pi}}{\sqrt{2}}(Y_1^{-1} Y_1^{+1} + Y_1^{+1} Y_1^{-1})  $ & $-\frac{3}{4\sqrt{\pi}} \sin\theta_1\sin\theta_2\cos\varphi$ \\ [1mm]
 1 & 1 & ${\perp      }$ & $ \frac{\sqrt{\pi}}{\sqrt{2}}(Y_1^{-1} Y_1^{+1} - Y_1^{+1} Y_1^{-1})  $ & $-i\frac{3}{4\sqrt{\pi}} \sin\theta_1\sin\theta_2\sin\varphi$ \\ [1mm] \hline
 1 & 2 & ${0          }$ & $ \sqrt{\pi} Y_1^0 Y_2^0  $                  & $\frac{\sqrt{15}}{4\sqrt{2\pi}}\cos\theta_1(3\cos^2\theta_2-1)$ \\ [1mm]
 1 & 2 & ${\parallel  }$ & $ \frac{\sqrt{\pi}}{\sqrt{2}}(Y_1^{-1} Y_2^{+1} + Y_1^{+1} Y_2^{-1})  $ & $\frac{3\sqrt{5}}{4\sqrt{\pi}}\sin\theta_1\sin\theta_2\cos\theta_2\cos\varphi$ \\ [1mm]
 1 & 2 & ${\perp      }$ & $ \frac{\sqrt{\pi}}{\sqrt{2}}(Y_1^{-1} Y_2^{+1} - Y_1^{+1} Y_2^{-1})  $ & $i\frac{3\sqrt{5}}{4\sqrt{\pi}}\sin\theta_1\sin\theta_2\cos\theta_2\sin\varphi$ \\ [1mm] \hline
 2 & 1 & ${0          }$ & $ \sqrt{\pi} Y_2^0 Y_1^0  $                  & $-\frac{\sqrt{15}}{4\sqrt{2\pi}}(3\cos^2\theta_1-1)\cos\theta_2$ \\ [1mm]
 2 & 1 & ${\parallel  }$ & $ \frac{\sqrt{\pi}}{\sqrt{2}}(Y_2^{-1} Y_1^{+1} + Y_2^{+1} Y_1^{-1})  $ & $-\frac{3\sqrt{5}}{4\sqrt{\pi}}\sin\theta_1\cos\theta_1\sin\theta_2\cos\varphi$ \\ [1mm]
 2 & 1 & ${\perp      }$ & $ \frac{\sqrt{\pi}}{\sqrt{2}}(Y_2^{-1} Y_1^{+1} - Y_2^{+1} Y_1^{-1})  $ & $-i\frac{3\sqrt{5}}{4\sqrt{\pi}}\sin\theta_1\cos\theta_1\sin\theta_2\sin\varphi$ \\ [1mm] \hline
 2 & 2 & ${0          }$ & $ \sqrt{\pi} Y_2^0 Y_2^0  $                  & $\frac{5}{8\sqrt{2\pi}}(3\cos^2\theta_1-1)(3\cos^2\theta_2-1)$ \\ [1mm]
 2 & 2 & ${\parallel_{1}}$ & $ \frac{\sqrt{\pi}}{\sqrt{2}}(Y_2^{-1} Y_2^{+1} + Y_2^{+1} Y_2^{-1})  $ & $\frac{15}{4\sqrt{\pi}}\sin\theta_1\cos\theta_1\sin\theta_2\cos\theta_2\cos\varphi$ \\ [1mm]
 2 & 2 & ${\perp_{1}}$ & $ \frac{\sqrt{\pi}}{\sqrt{2}}(Y_2^{-1} Y_2^{+1} - Y_2^{+1} Y_2^{-1})  $ & $i\frac{15}{4\sqrt{\pi}}\sin\theta_1\cos\theta_1\sin\theta_2\cos\theta_2\sin\varphi$ \\ [1mm]
 2 & 2 & ${\parallel_{2}}$ & $ \frac{\sqrt{\pi}}{\sqrt{2}}(Y_2^{-2} Y_2^{+2} + Y_2^{+2} Y_2^{-2})  $ & $\frac{15}{16\sqrt{\pi}}\sin^2\theta_1\sin^2\theta_2\cos(2\varphi)$ \\ [1mm]
 2 & 2 & ${\perp_{2}}$ & $ \frac{\sqrt{\pi}}{\sqrt{2}}(Y_2^{-2} Y_2^{+2} - Y_2^{+2} Y_2^{-2})  $ & $i\frac{15}{16\sqrt{\pi}}\sin^2\theta_1\sin^2\theta_2\sin(2\varphi)$ \\ [1mm] \hline
\end{tabular}
\label{tab:ang_amplitudes}
\end{center}
\end{table}
\section{Scalar $K\pi$ mass-dependent amplitude}
\label{sec:M0}
The variation of the phase with $m(K\pi)$ in the nominal model used for the scalar $K\pi$ mass-dependent amplitude is
taken from Ref.~\cite{PhysRevD.93.074025}. The modulus line-shape is parameterised with a polynomial
expansion as follows
\begin{equation}
|\mathcal{M}_0(m)|=1+\sum_{i=1}^4c_i\, T_{i}(X(m)),
\end{equation}
where $X(m)=(m-1175\MeVcc)/425\MeVcc$ $X(m)\in[-1,1]$, and $T_i(x)$ are the Chebyshev polynomials
defined as
\begin{equation}
\begin{tabular}{lll}
$T_{0}(x) = 1$, & $T_{1}(x) = x$, & $T_{2}(x) = 2x^{2}-1$, \\
$T_{3}(x) = 4x^{3}-3x$, & $T_{4}(x) = 8x^{4}-8x^{2}+1$. & \\
\end{tabular}
\end{equation}
This parameterisation is chosen to minimise parameter correlations.
The values of the $c_i$ coefficients retrieved from the decay-time-dependent fit are given in
\tabref{tab:M0_nominal}. The coefficients decrease with the order of the
polynomial term. The expansion is truncated at fourth order since adding an extra
term would not significantly affect the result and the size of the fifth coefficient is of
the order of its statistical uncertainty.
\begin{table}
\centering
\caption{Parameters used in the nominal model for the scalar $K\pi$ mass-dependent amplitude. The correlations among them are found to be small, the largest ones been of the order of $50\%$.}
\begin{tabular}{|cc|}
\hline
Parameter & Value \\
\hline
$c_{1}$ & $-0.287 \pm 0.020$ \\
$c_{2}$ & $-0.180 \pm 0.020$ \\
$c_{3}$ & $-0.106 \pm 0.016$ \\
$c_{4}$ & $-0.066 \pm 0.016$ \\
\hline
\end{tabular}
\label{tab:M0_nominal}
\end{table}
\begin{figure}[t]
\center
\includegraphics[width=0.48\textwidth]{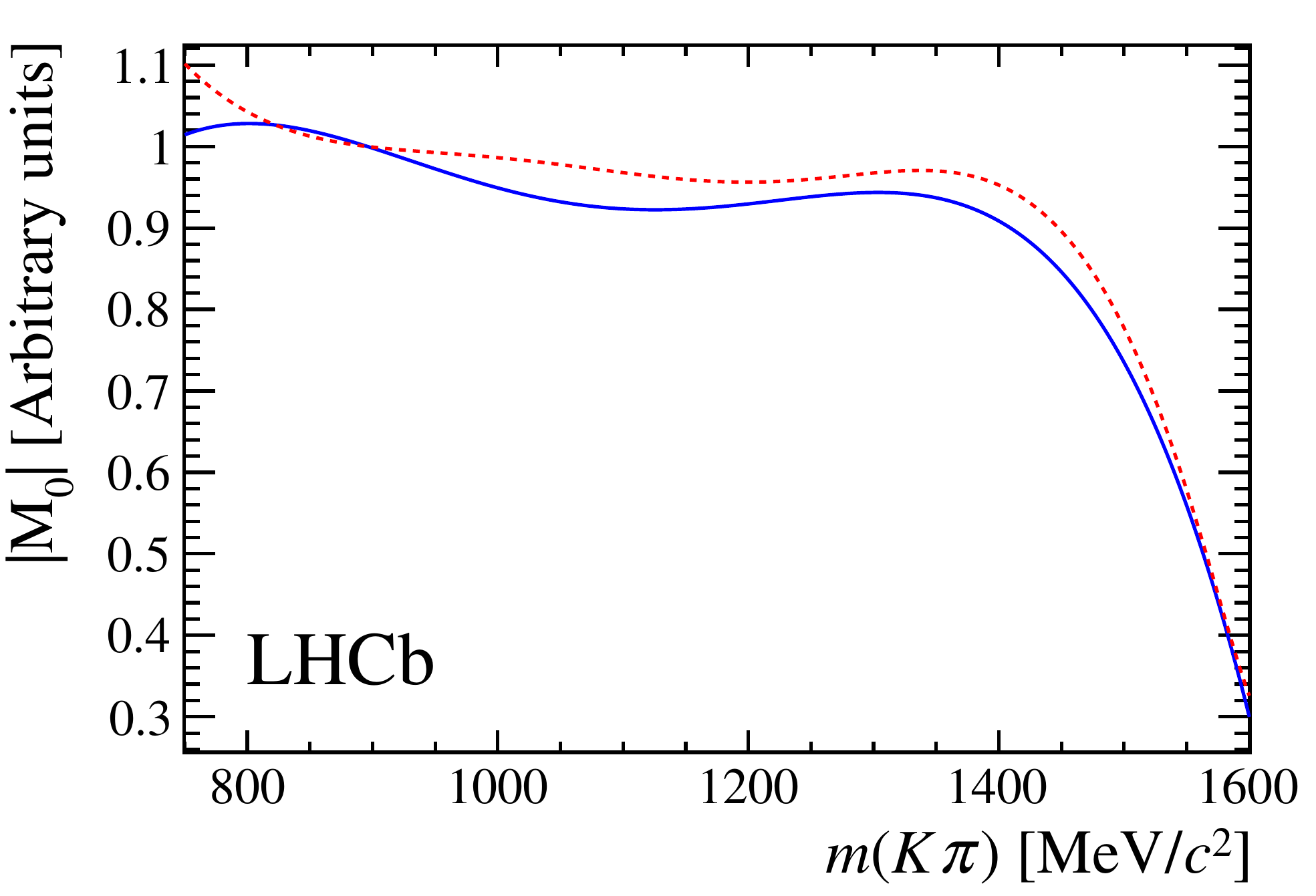}
\includegraphics[width=0.48\textwidth]{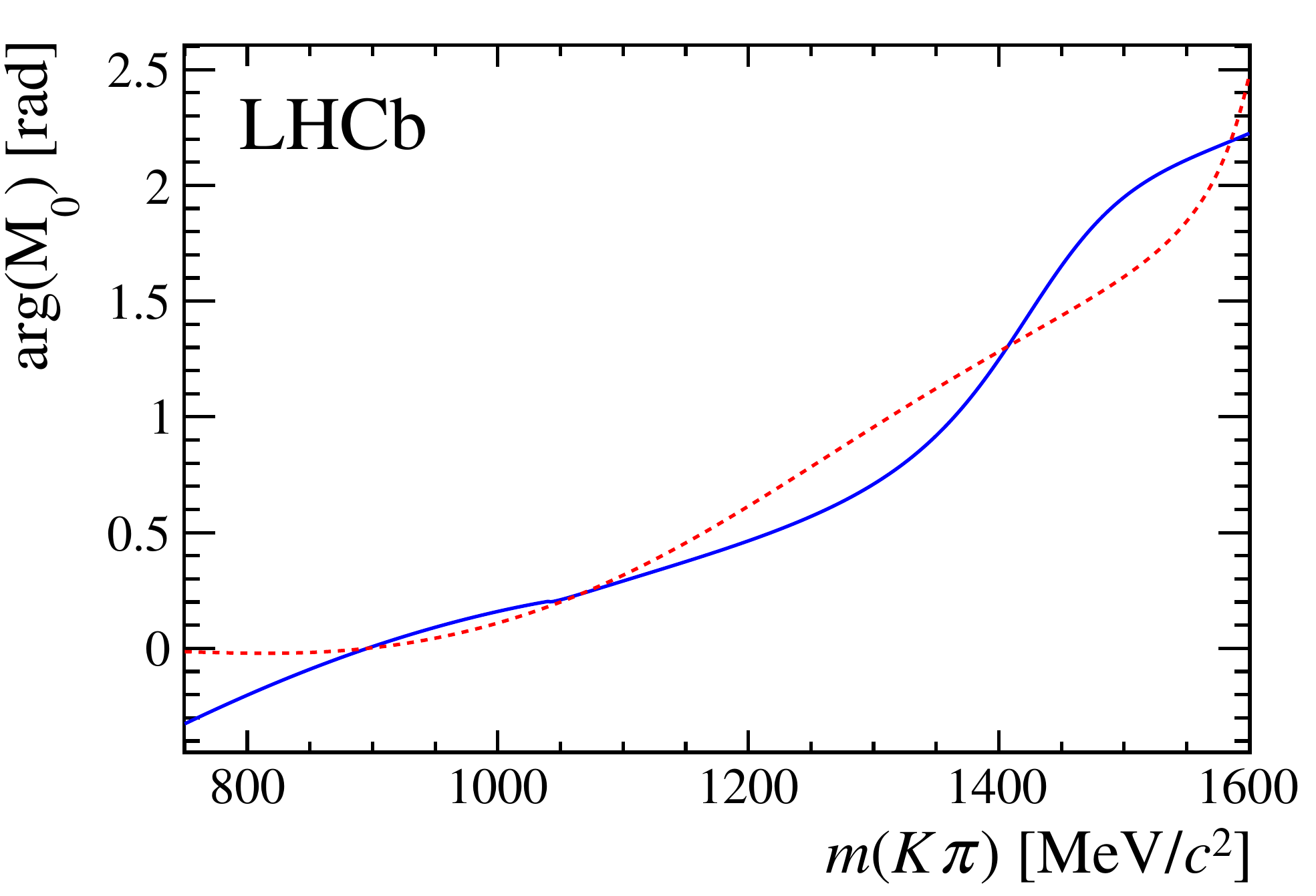}
\caption{Line-shapes of the (left) modulus and (right) phase of the scalar $K\pi$ mass-dependent
amplitude. The nominal model is shown with a solid blue line and the model-independent
parameterisation, used in systematic studies, is shown with a dashed red line.}
\label{fig:scalarmassamplitude}
\end{figure}
When computing systematic uncertainties, the scalar $K\pi$ mass-dependent amplitude is
parameterised using a model-independent (MI) approach as follows
\begin{equation}
\mathcal{M}_{0}^{MI}(m)=\left[1+\sum_{i=1}^{4}\alpha_{i}\, T_{i}(X(m))\right]+i
\,\left[\sum_{j=0}^{4}\beta_{j}\, T_{j}(X(m))\right].
\end{equation}
The coefficients measured in the decay-time-dependent fit for this case are given in \tabref{tab:M0_MI}.
\begin{table}
\centering
\caption{Coefficients used in the model-independent parameterisation of the scalar $K\pi$ mass-dependent amplitude.}
\begin{tabular}{|cc|cc|}
\hline
Parameter & Value & Parameter & Value \\
\hline
$\alpha_{1}$ & $-0.854  \pm  0.038$  & $\beta_{0}$ & $\phantom{-}0.278  \pm  0.038$ \\
$\alpha_{2}$ & $-0.381  \pm  0.040$  & $\beta_{1}$ & $\phantom{-}0.817  \pm 0.079$ \\
$\alpha_{3}$ & $-0.105   \pm  0.032$ & $\beta_{2}$ & $-0.206  \pm  0.082$ \\
$\alpha_{4}$ & $\phantom{-}0.046 \pm  0.027$ & $\beta_{3}$ & $-0.367  \pm  0.053$ \\
             &                         & $\beta_{4}$ & $-0.115  \pm  0.040$ \\
\hline
\end{tabular}
\label{tab:M0_MI}
\end{table}
The line-shapes of the two scalar mass amplitude models are shown in
\figref{fig:scalarmassamplitude}. Both approaches are found to be qualitatively compatible with
each other.

\addcontentsline{toc}{section}{References}
\setboolean{inbibliography}{true}
\bibliographystyle{LHCb}
\bibliography{main,LHCb-PAPER,LHCb-CONF,LHCb-DP,LHCb-TDR}

\newpage

\newpage
\centerline{\large\bf LHCb collaboration}
\begin{flushleft}
\small
R.~Aaij$^{40}$,
B.~Adeva$^{39}$,
M.~Adinolfi$^{48}$,
Z.~Ajaltouni$^{5}$,
S.~Akar$^{59}$,
J.~Albrecht$^{10}$,
F.~Alessio$^{40}$,
M.~Alexander$^{53}$,
A.~Alfonso~Albero$^{38}$,
S.~Ali$^{43}$
% ...
{\footnotesize \it
% ...
$ ^{5}$Clermont Universit{\'e}, Universit{\'e} Blaise Pascal, CNRS/IN2P3, LPC, Clermont-Ferrand, France\\
% ...
$ ^{10}$Fakult{\"a}t Physik, Technische Universit{\"a}t Dortmund, Dortmund, Germany\\
% ...
$ ^{38}$ICCUB, Universitat de Barcelona, Barcelona, Spain\\
$ ^{39}$Instituto Galego de F{\'\i}sica de Altas Enerx{\'\i}as (IGFAE), Universidade de Santiago de Compostela, Santiago de Compostela, Spain\\
$ ^{40}$European Organization for Nuclear Research (CERN), Geneva, Switzerland\\
% ...
$ ^{43}$Nikhef National Institute for Subatomic Physics, Amsterdam, The Netherlands\\
% ...
$ ^{48}$H.H. Wills Physics Laboratory, University of Bristol, Bristol, United Kingdom\\
% ...
$ ^{53}$School of Physics and Astronomy, University of Glasgow, Glasgow, United Kingdom\\
% ...
$ ^{59}$University of Cincinnati, Cincinnati, OH, United States\\
% ...
}
\end{flushleft}

\end{document}